%% file: main.tex
\documentclass[11pt]{article}

\usepackage[margin=1in]{geometry}
\usepackage{amsmath}
\usepackage{amsfonts}
\usepackage{amssymb}
\usepackage{amsthm}
\usepackage{graphics}
\usepackage{graphicx}
\usepackage{physics}
\usepackage[linesnumbered,ruled,vlined]{algorithm2e}
\usepackage[dvipsnames]{xcolor}
\usepackage[inline,shortlabels]{enumitem}
\usepackage{authblk}
\usepackage{bbm}
\usepackage{appendix}
\usepackage{tcolorbox}
\usepackage[numbers,square]{natbib}
\usepackage{microtype}
\usepackage[normalem]{ulem}
\usepackage{booktabs}
\usepackage{enumitem}
\usepackage[T1]{fontenc}
\usepackage[utf8]{inputenc}
\usepackage[american]{babel}
\usepackage{todonotes}

\usepackage[colorlinks=true,linktoc=page]{hyperref}
\hypersetup{colorlinks,
  linkcolor=blue,
  citecolor=ForestGreen,  
  urlcolor=Mahogany,
}
\usepackage{cleveref}

\newtheorem{theorem}{Theorem}
\newtheorem{lemma}[theorem]{Lemma}

\newtheorem{definition}[theorem]{Definition}

\title{A Fast and Accurate Algorithm for the $p\to q$ Norm of Sparse Non-Negative Matrices\thanks{This research was partially supported by
    ONR Award N00014-24-12611.}}
\author{Marco Pirazzini\thanks{Email: \texttt{marco.pirazzini@yale.edu}} }
\author{Daniel A. Spielman\thanks{Email: \texttt{daniel.spielman@yale.edu}}}
\affil{Yale University}
\date{November 6, 2026}

\begin{document}

\include{sgt_defs}

\def\Otilde{\widetilde{O}}

\maketitle

\begin{abstract}
We present an algorithm that computes the $p \to q$ norm of a non-negative matrix with $s$ nonzero entries to accuracy $\varepsilon$ in time $\Otilde(s^{3/2} \log (1/\varepsilon))$, for any $p \ge q \ge 1$.
The fastest algorithm previously designed for this problem has running time $\Otilde(s / \varepsilon)$
  (Objois and Vladu, 2026),
  which is slower when high accuracy solutions are desired.
  
Our algorithm employs a surprising application of Laplacian solvers inside an Interior Point Method (IPM).
We formulate the problem as a linear optimization problem over a product of power cones,
and show that the Newton updates of the IPM can be computed by solving a Laplacian system in a Schur complement after subtracting a rank-2 matrix.

We also discuss an implementation of our algorithm, showing that it is practical and it outperforms optimized solvers such as MOSEK by factors of up to $20\times$ on a variety of tasks.

\end{abstract}

\thispagestyle{empty}
\clearpage
\pagenumbering{arabic}

\input{sections/01_intro}

\input{sections/02_preliminaries}

\input{sections/03_conic}

\input{sections/04_IPMsec}

\input{sections/05_newton}

\input{sections/06_experiments}

\section*{AI Methodology}

We discovered the conic formulation of the problem and started to investigate using interior point methods to perform the optimization. ChatGPT-6 Pro suggested the acceleration of the interior point method by the use of Laplacian solvers. We substantially simplified the proof and exposition, with some assistance from ChatGPT-6 Pro. All experiments were designed by the authors, and we used Claude Code for implementation. We used ChatGPT and Claude to check all the calculations in the paper and the implementations.

The paper was entirely written by the authors, who verified the correctness and originality of all content, including code and references.

\bibliographystyle{alpha}
\bibliography{references}

\appendix

\input{sections/A_IPMapp}

\input{sections/B_derivatives}

\input{sections/C_experiments_app}

\end{document}

%% file: sgt_defs.tex
\def\field#1{\mathbb{F}_{#1}}

\def\biE{\overrightarrow{\mathrm{E}}}

\def\Reff{\mathrm{R_{eff}}}
\def\reff{\mathrm{Reff}}
\def\Reffhat{\mathrm{\widehat{R}_{eff}}}
\def\Ceff{\mathrm{C_{eff}}}

\def\Stretch{\mathrm{Stretch}}

\def\calSO{\mathcal{SO}}

\def\cdv#1{\mathsf{cdv} (#1)}

\def\st#1#2{\mathsf{st}_{#1} (#2)}

\def\sp#1{\mathsf{sp} (#1)}
\def\Sp#1{\mathsf{sp}_{#1}}

\def\floor#1{\left\lfloor #1 \right\rfloor}
\def\ceil#1{\left\lceil #1 \right\rceil}

\def\dim#1{\mathrm{dim} (#1)}

\def\sgn#1{\mathsf{sgn} (#1)}

\def\aut#1{\mathsf{aut}(#1)}

\def\dist#1#2{\mathrm{dist} (#1, #2)}

\def\trace#1{\mathrm{Tr} \left(#1 \right)}
\def\trop{\mathrm{Tr }}


\def\calC{\mathcal{C}}


\def\calS{\mathcal{S}}
\def\intersect{\cap}
\def\form#1#2{\left\langle #1, #2\right\rangle}



\def\ceil#1{\left\lceil #1 \right\rceil}
\def\vol#1{\mathrm{vol} (#1)}
\def\bdry#1{\partial (#1)}


\def\abs#1{\left|#1  \right|}

\def\norm#1{\left\| #1 \right\|}
\def\smallnorm#1{\| #1 \|}

\def\nucnorm#1{\left\| #1 \right\|_{*}}

\def\vs#1#2#3{#1_{#2},\ldots , #1_{#3}}

\newcommand{\energy}[1]{\mathcal{E}\left(#1 \right)}

\newcommand\zzero{\boldsymbol{0}}
\newcommand\oone{\boldsymbol{1}}

\def\calA{\mathcal{A}}
\def\calB{\mathcal{B}}
\def\calC{\mathcal{C}}
\def\calD{\mathcal{D}}
\def\calE{\mathcal{E}}
\def\calF{\mathcal{F}}
\def\calG{\mathcal{G}}
\def\calH{\mathcal{H}}
\def\calI{\mathcal{I}}
\def\calJ{\mathcal{J}}
\def\calK{\mathcal{K}}
\def\calL{\mathcal{L}}
\def\calM{\mathcal{M}}
\def\calN{\mathcal{N}}
\def\calO{\mathcal{O}}
\def\calP{\mathcal{P}}
\def\calQ{\mathcal{Q}}
\def\calR{\mathcal{R}}
\def\calS{\mathcal{S}}
\def\calT{\mathcal{T}}
\def\calU{\mathcal{U}}
\def\calV{\mathcal{V}}
\def\calW{\mathcal{W}}
\def\calX{\mathcal{X}}
\def\calY{\mathcal{Y}}
\def\calZ{\mathcal{Z}}

\renewcommand\AA{\boldsymbol{\mathit{A}}}
\newcommand\BB{\boldsymbol{\mathit{B}}}
\newcommand\CC{\boldsymbol{\mathit{C}}}
\newcommand\DD{\boldsymbol{\mathit{D}}}
\newcommand\EE{\boldsymbol{\mathit{E}}}
\newcommand\FF{\boldsymbol{\mathit{F}}}
\newcommand\GG{\boldsymbol{\mathit{G}}}
\newcommand\HH{\boldsymbol{\mathit{H}}}
\newcommand\II{\boldsymbol{\mathit{I}}}
\newcommand\JJ{\boldsymbol{\mathit{J}}}
\newcommand\KK{\boldsymbol{\mathit{K}}}
\newcommand\LL{\boldsymbol{\mathit{L}}}
\newcommand\MM{\boldsymbol{\mathit{M}}}
\newcommand\NN{\boldsymbol{\mathit{N}}}
\newcommand\OO{\boldsymbol{\mathit{O}}}
\newcommand\PP{\boldsymbol{\mathit{P}}}
\newcommand\QQ{\boldsymbol{\mathit{Q}}}
\newcommand\RR{\boldsymbol{\mathit{R}}}
\renewcommand\SS{\boldsymbol{\mathit{S}}}
\newcommand\TT{\boldsymbol{\mathit{T}}}
\newcommand\UU{\boldsymbol{\mathit{U}}}
\newcommand\VV{\boldsymbol{\mathit{V}}}
\newcommand\WW{\boldsymbol{\mathit{W}}}
\newcommand\XX{\boldsymbol{\mathit{X}}}
\newcommand\YY{\boldsymbol{\mathit{Y}}}
\newcommand\ZZ{\boldsymbol{\mathit{Z}}}

\def\aa{\pmb{\mathit{a}}}
\newcommand\bb{\boldsymbol{\mathit{b}}}
\newcommand\cc{\boldsymbol{\mathit{c}}}
\renewcommand\dd{\boldsymbol{\mathit{d}}}
\newcommand\ee{\boldsymbol{\mathit{e}}}
\newcommand\ff{\boldsymbol{\mathit{f}}}
\renewcommand\gg{\boldsymbol{\mathit{g}}}
\newcommand\hh{\boldsymbol{\mathit{h}}}
\newcommand\ii{\boldsymbol{\mathit{i}}}
\newcommand\jj{\boldsymbol{\mathit{j}}}
\newcommand\kk{\boldsymbol{\mathit{k}}}
\renewcommand\ll{\boldsymbol{\mathit{l}}}

\newcommand\pp{\boldsymbol{\mathit{p}}}

\renewcommand\qq{\boldsymbol{\mathit{q}}}

\newcommand\bs{\boldsymbol{\mathit{s}}}
\newcommand\nn{\boldsymbol{\mathit{n}}}
\newcommand\rr{\boldsymbol{\mathit{r}}}
\renewcommand\ss{\boldsymbol{\mathit{s}}}
\newcommand\uu{\boldsymbol{\mathit{u}}}
\def\vv{\boldsymbol{\mathit{v}}}
\newcommand\ww{\boldsymbol{\mathit{w}}}
\newcommand\xx{\boldsymbol{\mathit{x}}}

\newcommand\xp{\boldsymbol{\mathit{x'}}}
\newcommand\yy{\boldsymbol{\mathit{y}}}
\newcommand\zz{\boldsymbol{\mathit{z}}}
\def\tt{\boldsymbol{\mathit{t}}}
\newcommand\mm{\boldsymbol{\mathit{m}}}

\newcommand\aatil{\boldsymbol{\widetilde{\mathit{a}}}}
\newcommand\bbtil{\boldsymbol{\widetilde{\mathit{b}}}}
\newcommand\cctil{\boldsymbol{\widetilde{\mathit{c}}}}
\newcommand\ddtil{\boldsymbol{\widetilde{\mathit{d}}}}
\newcommand\eetil{\boldsymbol{\widetilde{\mathit{e}}}}
\newcommand\fftil{\boldsymbol{\widetilde{\mathit{f}}}}
\newcommand\ggtil{\boldsymbol{\widetilde{\mathit{g}}}}
\newcommand\hhtil{\boldsymbol{\widetilde{\mathit{h}}}}
\newcommand\iitil{\boldsymbol{\widetilde{\mathit{i}}}}
\newcommand\jjtil{\boldsymbol{\widetilde{\mathit{j}}}}
\newcommand\kktil{\boldsymbol{\widetilde{\mathit{k}}}}
\newcommand\lltil{\boldsymbol{\widetilde{\mathit{l}}}}
\newcommand\mmtil{\boldsymbol{\widetilde{\mathit{m}}}}
\newcommand\nntil{\boldsymbol{\widetilde{\mathit{n}}}}
\newcommand\ootil{\boldsymbol{\widetilde{\mathit{o}}}}
\newcommand\pptil{\boldsymbol{\widetilde{\mathit{p}}}}
\newcommand\qqtil{\boldsymbol{\widetilde{\mathit{q}}}}
\newcommand\rrtil{\boldsymbol{\widetilde{\mathit{r}}}}
\newcommand\sstil{\boldsymbol{\widetilde{\mathit{s}}}}
\newcommand\tttil{\boldsymbol{\widetilde{\mathit{t}}}}
\newcommand\uutil{\boldsymbol{\widetilde{\mathit{u}}}}
\newcommand\vvtil{\boldsymbol{\widetilde{\mathit{v}}}}
\newcommand\wwtil{\boldsymbol{\widetilde{\mathit{w}}}}
\newcommand\xxtil{\boldsymbol{\widetilde{\mathit{x}}}}
\newcommand\yytil{\boldsymbol{\widetilde{\mathit{y}}}}
\newcommand\zztil{\boldsymbol{\widetilde{\mathit{z}}}}

\newcommand\aahat{\boldsymbol{\widehat{\mathit{a}}}}
\newcommand\bbhat{\boldsymbol{\widehat{\mathit{b}}}}
\newcommand\cchat{\boldsymbol{\widehat{\mathit{c}}}}
\newcommand\ddhat{\boldsymbol{\widehat{\mathit{d}}}}
\newcommand\eehat{\boldsymbol{\widehat{\mathit{e}}}}
\newcommand\ffhat{\boldsymbol{\widehat{\mathit{f}}}}
\newcommand\gghat{\boldsymbol{\widehat{\mathit{g}}}}
\newcommand\hhhat{\boldsymbol{\widehat{\mathit{h}}}}
\newcommand\iihat{\boldsymbol{\widehat{\mathit{i}}}}
\newcommand\jjhat{\boldsymbol{\widehat{\mathit{j}}}}
\newcommand\kkhat{\boldsymbol{\widehat{\mathit{k}}}}
\newcommand\llhat{\boldsymbol{\widehat{\mathit{l}}}}
\newcommand\mmhat{\boldsymbol{\widehat{\mathit{m}}}}
\newcommand\nnhat{\boldsymbol{\widehat{\mathit{n}}}}
\newcommand\oohat{\boldsymbol{\widehat{\mathit{o}}}}
\newcommand\pphat{\boldsymbol{\widehat{\mathit{p}}}}
\newcommand\qqhat{\boldsymbol{\widehat{\mathit{q}}}}
\newcommand\rrhat{\boldsymbol{\widehat{\mathit{r}}}}
\newcommand\sshat{\boldsymbol{\widehat{\mathit{s}}}}
\newcommand\tthat{\boldsymbol{\widehat{\mathit{t}}}}
\newcommand\uuhat{\boldsymbol{\widehat{\mathit{u}}}}
\newcommand\vvhat{\boldsymbol{\widehat{\mathit{v}}}}
\newcommand\wwhat{\boldsymbol{\widehat{\mathit{w}}}}
\newcommand\xxhat{\boldsymbol{\widehat{\mathit{x}}}}
\newcommand\yyhat{\boldsymbol{\widehat{\mathit{y}}}}
\newcommand\zzhat{\boldsymbol{\widehat{\mathit{z}}}}

\newcommand\AAhat{\boldsymbol{\widehat{\mathit{A}}}}
\newcommand\BBhat{\boldsymbol{\widehat{\mathit{B}}}}
\newcommand\CChat{\boldsymbol{\widehat{\mathit{C}}}}
\newcommand\DDhat{\boldsymbol{\widehat{\mathit{D}}}}
\newcommand\EEhat{\boldsymbol{\widehat{\mathit{E}}}}
\newcommand\FFhat{\boldsymbol{\widehat{\mathit{F}}}}
\newcommand\GGhat{\boldsymbol{\widehat{\mathit{G}}}}
\newcommand\HHhat{\boldsymbol{\widehat{\mathit{H}}}}
\newcommand\IIhat{\boldsymbol{\widehat{\mathit{I}}}}
\newcommand\JJhat{\boldsymbol{\widehat{\mathit{J}}}}
\newcommand\KKhat{\boldsymbol{\widehat{\mathit{K}}}}
\newcommand\LLhat{\boldsymbol{\widehat{\mathit{L}}}}
\newcommand\MMhat{\boldsymbol{\widehat{\mathit{M}}}}
\newcommand\NNhat{\boldsymbol{\widehat{\mathit{N}}}}
\newcommand\OOhat{\boldsymbol{\widehat{\mathit{O}}}}
\newcommand\PPhat{\boldsymbol{\widehat{\mathit{P}}}}
\newcommand\QQhat{\boldsymbol{\widehat{\mathit{Q}}}}
\newcommand\RRhat{\boldsymbol{\widehat{\mathit{R}}}}
\newcommand\SShat{\boldsymbol{\widehat{\mathit{S}}}}
\newcommand\TThat{\boldsymbol{\widehat{\mathit{T}}}}
\newcommand\UUhat{\boldsymbol{\widehat{\mathit{U}}}}
\newcommand\VVhat{\boldsymbol{\widehat{\mathit{V}}}}
\newcommand\WWhat{\boldsymbol{\widehat{\mathit{W}}}}
\newcommand\XXhat{\boldsymbol{\widehat{\mathit{X}}}}
\newcommand\YYhat{\boldsymbol{\widehat{\mathit{Y}}}}
\newcommand\ZZhat{\boldsymbol{\widehat{\mathit{Z}}}}

\newcommand\AAtil{\boldsymbol{\widetilde{\mathit{A}}}}
\newcommand\BBtil{\boldsymbol{\widetilde{\mathit{B}}}}
\newcommand\CCtil{\boldsymbol{\widetilde{\mathit{C}}}}
\newcommand\DDtil{\boldsymbol{\widetilde{\mathit{D}}}}
\newcommand\EEtil{\boldsymbol{\widetilde{\mathit{E}}}}
\newcommand\FFtil{\boldsymbol{\widetilde{\mathit{F}}}}
\newcommand\GGtil{\boldsymbol{\widetilde{\mathit{G}}}}
\newcommand\HHtil{\boldsymbol{\widetilde{\mathit{H}}}}
\newcommand\IItil{\boldsymbol{\widetilde{\mathit{I}}}}
\newcommand\JJtil{\boldsymbol{\widetilde{\mathit{J}}}}
\newcommand\KKtil{\boldsymbol{\widetilde{\mathit{K}}}}
\newcommand\LLtil{\boldsymbol{\widetilde{\mathit{L}}}}
\newcommand\MMtil{\boldsymbol{\widetilde{\mathit{M}}}}
\newcommand\NNtil{\boldsymbol{\widetilde{\mathit{N}}}}
\newcommand\OOtil{\boldsymbol{\widetilde{\mathit{O}}}}
\newcommand\PPtil{\boldsymbol{\widetilde{\mathit{P}}}}
\newcommand\QQtil{\boldsymbol{\widetilde{\mathit{Q}}}}
\newcommand\RRtil{\boldsymbol{\widetilde{\mathit{R}}}}
\newcommand\SStil{\boldsymbol{\widetilde{\mathit{S}}}}
\newcommand\TTtil{\boldsymbol{\widetilde{\mathit{T}}}}
\newcommand\UUtil{\boldsymbol{\widetilde{\mathit{U}}}}
\newcommand\VVtil{\boldsymbol{\widetilde{\mathit{V}}}}
\newcommand\WWtil{\boldsymbol{\widetilde{\mathit{W}}}}
\newcommand\XXtil{\boldsymbol{\widetilde{\mathit{X}}}}
\newcommand\YYtil{\boldsymbol{\widetilde{\mathit{Y}}}}
\newcommand\ZZtil{\boldsymbol{\widetilde{\mathit{Z}}}}

\newcommand\Atil{\widetilde{\mathit{A}}}
\newcommand\Btil{\widetilde{\mathit{B}}}
\newcommand\Ctil{\widetilde{\mathit{C}}}
\newcommand\Dtil{\widetilde{\mathit{D}}}
\newcommand\Etil{\widetilde{\mathit{E}}}
\newcommand\Ftil{\widetilde{\mathit{F}}}
\newcommand\Gtil{\widetilde{\mathit{G}}}
\newcommand\Htil{\widetilde{\mathit{H}}}
\newcommand\Itil{\widetilde{\mathit{I}}}
\newcommand\Jtil{\widetilde{\mathit{J}}}
\newcommand\Ktil{\widetilde{\mathit{K}}}
\newcommand\Ltil{\widetilde{\mathit{L}}}
\newcommand\Mtil{\widetilde{\mathit{M}}}
\newcommand\Ntil{\widetilde{\mathit{N}}}
\newcommand\Otil{\widetilde{\mathit{O}}}
\newcommand\Ptil{\widetilde{\mathit{P}}}
\newcommand\Qtil{\widetilde{\mathit{Q}}}
\newcommand\Rtil{\widetilde{\mathit{R}}}
\newcommand\Stil{\widetilde{\mathit{S}}}
\newcommand\Ttil{\widetilde{\mathit{T}}}
\newcommand\Util{\widetilde{\mathit{U}}}
\newcommand\Vtil{\widetilde{\mathit{V}}}
\newcommand\Wtil{\widetilde{\mathit{W}}}
\newcommand\Xtil{\widetilde{\mathit{X}}}
\newcommand\Ytil{\widetilde{\mathit{Y}}}
\newcommand\Ztil{\widetilde{\mathit{Z}}}

\newcommand\aalpha{\boldsymbol{\alpha}}
\newcommand\bbeta{\boldsymbol{\beta}}
\newcommand\ggamma{\boldsymbol{\gamma}}
\newcommand\ddelta{\boldsymbol{\delta}}
\newcommand\eepsilon{\boldsymbol{\epsilon}}
\newcommand{\zzeta}{\boldsymbol{\zeta}}
\newcommand{\eeta}{\boldsymbol{\eta}}
\newcommand{\ttheta}{\boldsymbol{\theta}}
\newcommand{\iiota}{\boldsymbol{\iota}}
\newcommand{\kkappa}{\boldsymbol{\kappa}}
\newcommand\llambda{\boldsymbol{\lambda}}
\newcommand\mmu{\boldsymbol{\mu}}
\newcommand\nnu{\boldsymbol{\nu}}
\newcommand{\xxi}{\boldsymbol{\xi}}
\newcommand\ppi{\boldsymbol{\pi}}
\newcommand{\rrho}{\boldsymbol{\rho}}
\newcommand\ssigma{\boldsymbol{\sigma}}
\newcommand\ttau{\boldsymbol{\tau}}
\newcommand\pphi{\boldsymbol{\phi}}
\newcommand{\cchi}{\boldsymbol{\chi}}
\newcommand\ppsi{\boldsymbol{\psi}}
\newcommand\oomega{\boldsymbol{\omega}}

\newcommand\GGamma{\boldsymbol{\Gamma}}
\newcommand{\DDelta}{\boldsymbol{\Delta}}
\newcommand{\TTheta}{\boldsymbol{\Theta}}
\newcommand{\LLambda}{\boldsymbol{\Lambda}}
\newcommand{\PPi}{\boldsymbol{\Pi}}
\newcommand{\SSigma}{\boldsymbol{\Sigma}}
\newcommand{\UUpsilon}{\boldsymbol{\Upsilon}}
\newcommand{\PPhi}{\boldsymbol{\Phi}}
\newcommand{\PPsi}{\boldsymbol{\Psi}}
\newcommand{\OOmega}{\boldsymbol{\Omega}}

\newcommand\Ggamma{\boldsymbol{\Gamma}}
\newcommand{\Ddelta}{\boldsymbol{\Delta}}
\newcommand{\Ttheta}{\boldsymbol{\Theta}}
\newcommand{\Llambda}{\boldsymbol{\Lambda}}
\newcommand{\Ppi}{\boldsymbol{\Pi}}
\newcommand{\Ssigma}{\boldsymbol{\Sigma}}
\newcommand{\Uupsilon}{\boldsymbol{\Upsilon}}
\newcommand{\Pphi}{\boldsymbol{\Phi}}
\newcommand{\Ppsi}{\boldsymbol{\Psi}}
\newcommand{\Oomega}{\boldsymbol{\Omega}}

\newcommand\xhat{\boldsymbol{\hat{\mathit{x}}}}
\newcommand\ahat{{\widehat{{a}}}}
\newcommand\bhat{{\widehat{{b}}}}
\newcommand\chat{{\widehat{{c}}}}
\newcommand\uhat{{\widehat{{u}}}}
\newcommand\vhat{{\widehat{{v}}}}
\newcommand\what{{\widehat{{w}}}}

\newcommand\Ghat{{\widehat{{G}}}}

\newcommand\thetahat{{\hat{\theta }}}

\newcommand\lazy{\WWtil}

\def\union{\cup}

\newdimen\pIR
\pIR= -131072sp
\newcommand\StevesR{{\rm I\kern\pIR R}}
\def\Reals#1{\StevesR^{#1}}

\def\pleq{\preccurlyeq}
\def\pgeq{\succcurlyeq}

\def\ple{\prec}
\def\pge{\succ}

\def\Span#1{\textbf{Span}\left(#1  \right)}
\def\bvec#1{\mathbf{#1}}


\def\prob#1#2{\mbox{Pr}_{#1}\left[ #2 \right]}
\def\pvec#1#2{\vec{\mbox{P}}^{#1}\left[ #2 \right]}
\def\expec#1#2{{\mathop{\mathbb{E}}}_{#1}\left[ #2 \right]}
\def\var#1{\mbox{\bf Var}\left[ #1 \right]}
\def\E{\mbox{{\bf E}}}

\def\symExpec{\mathbb{E}}

\def\defeq{\stackrel{\mathrm{def}}{=}}
\def\setof#1{\left\{#1  \right\}}
\def\mand{\mbox{ and }}
\def\mor{\mbox{ or }}

\def\sizeof#1{\left|#1  \right|}

\newsavebox{\Codefig}

\newenvironment{codefig}[1]
{
\sbox{\Codefig}{\centering\epsfig{figure=figs/#1,width=4in}}%
\noindent
\begin{minipage}{.3\linewidth}
\begin{alltt}}
{\end{alltt}
\end{minipage}\hfill
\begin{minipage}{.6\linewidth}
\usebox{\Codefig}
\end{minipage}
}

\newenvironment{codeimg}[1]
{
\sbox{\Codefig}{\centering\epsfig{figure=figs/#1,width=4in}}%
\begin{center}
\begin{minipage}{.7\linewidth}
\usebox{\Codefig}
\begin{alltt}}
{\end{alltt}
\end{minipage}
\end{center}
}

\newenvironment{codeimg_label}[2]
{
\sbox{\Codefig}{\centering\epsfig{figure=figs/#1,width=4in}}%
\begin{center}
\begin{minipage}{.7\linewidth}
\usebox{\Codefig}
\begin{alltt}}
{\end{alltt}
\end{minipage}
\end{center}
}

\newenvironment{codefigRev}[1]
{
\sbox{\Codefig}{\centering\epsfig{figure=figs/#1,width=4in}}%
\noindent
\begin{minipage}{.3\linewidth}
\begin{alltt}}
{\end{alltt}
\end{minipage}\hfill
\begin{minipage}{.6\linewidth}
\usebox{\Codefig}
\end{minipage}
}

\newenvironment{codefig2}[1]
{
\sbox{\Codefig}{\centering\epsfig{figure=figs/#1,height=2.5in}}%
\noindent
\begin{minipage}{.45\linewidth}
\begin{alltt}}
{\end{alltt}
\end{minipage}\hfill
\begin{minipage}{.45\linewidth}
\usebox{\Codefig}
\end{minipage}
}

\newenvironment{codefig3}[1]
{
\sbox{\Codefig}{\centering\epsfig{figure=figs/#1,width=.45\linewidth}}%
\noindent
\begin{minipage}{.45\linewidth}
\begin{alltt}}
{\end{alltt}
\end{minipage}\hfill
\begin{minipage}{.45\linewidth}
\usebox{\Codefig}
\end{minipage}
}

\def\vecE{\vec{E}}

\def\id{\mathit{id}}

\def\lprod{{\ooalign{\hfil\raise .00ex\hbox{\scriptsize L}\hfil\crcr\mathhexbox20D}}}

\def\acos{\operatorname{acos}}
\def\acosh{\operatorname{acosh}}
\def\asinh{\operatorname{asinh}}

\def\csum#1#2{#1 \{ #2 \}}  
  
\def\match#1{\mu_{x}\left[ #1 \right]}

\newcommand{\verdiere}{Verdi{\`e}re }
\newcommand{\lovasz}{Lov{\`a}sz }
\newcommand{\poincare}{Poincar{\'e}}
\newcommand{\erdos}{Erd{\"o}s}
\newcommand{\renyi}{R{\'e}nyi}

%% file: sections/01_intro.tex
\section{Introduction}

The $p \to q$ norm of a matrix $\AA$ is 
\[
  \norm{\AA}_{p \to q} \defeq 
  \max_{\xx : \norm{\xx}_p = 1} \norm{\AA \xx}_q.
\]
When $p=q=2$, this quantity becomes the leading singular value of $\AA$ and $\xx$ becomes the singular vector.

The success of singular values and singular vectors has inspired quests to generalize them to solve new problems or improve existing solutions.
Their generalization for $p \ge q \ge 1$ has many potential applications, but 
the challenges in computing these has made them difficult to realize.

For general matrices, we only know efficient algorithms for computing these matrix norms when $p = 1$, $q = \infty$ or when $p = q \in \{1,2,\infty\}$. 
The \textit{hypercontractive} regime corresponds to $1 \le p < q$, while $p \ge q \ge 1$ is the \textit{non-hypercontractive} regime. For the latter, Bhaskara and Vijayaraghavan~\cite{bhaskara2011approximating} showed that it is NP-hard to approximate $\norm{\AA}_{p \to q}$ to within any constant factor for $1 < q \le p < 2$ or $2 < q \le p < \infty$, without any assumptions about $\AA$. For $p \ge 2 \ge q$, Bhattiprolu et al. \cite{bhattiprolu2019approximability} present a constant factor approximation algorithm using semidefinite programming. The hypercontractive regime has also received a lot of attention both for hardness of approximation results and specialized approximation algorithms \cite{barak2012hypercontractivity, bhattiprolu2018inapproximability, guth2025estimating, hopkins2026algorithms}.

For non-negative matrices $\AA$, \cite{bhaskara2011approximating} describe a polynomial-time approximation algorithm for the case $p \ge q \ge 1$.
The fastest algorithm for sparse matrices with $s$ non-zero entries and low accuracy is that of Objois and Vladu~\cite{objois2026approximating}, which obtains an $\varepsilon$ approximation in time $\Otilde(s / \epsilon)$.
To obtain running time depending on $\log(1/\varepsilon)$, one can apply standard convex programming techniques to the convex formulation of~\cite[Remark 3.4]{steinberg2005computation}.
The cutting planes method of Jiang, Lee, Song, and Wong~\cite{jiang2020improved} provides the asymptotically fastest high-accuracy algorithm for dense matrices based on this approach.

Computing the $p \to q$ norm of non-negative matrices for $p \ge q \ge 1$ has several applications, including oblivious routing schemes \cite{englert2009oblivious, bhaskara2011approximating, objois2026approximating}, robust optimization \cite{steinberg2005computation}, computing certifiable stopping criteria for Glauber dynamics \cite{dyer2009sampling}, or computing certificates for bounding errors in vector-valued predictions in neural networks \cite{combettes2020lipschitz}.

The following theorem encapsulates the guarantees of our algorithm, which is the fastest for high accuracy solutions in sparse non-negative matrices.

\begin{theorem}\label{thm::main}
On input a non-negative matrix $\AA \in \mathbb{R}^{m \times n}$ with $s$ non-zero entries, two numbers $p \ge q \ge 1$, and a $\varepsilon > 0$,
Algorithm \ref{alg::IPM} computes a $(1-\varepsilon)$ multiplicative approximation to $\|\AA\|_{p \to q}$ in expected time $\widetilde{O}(s^{\frac{3}{2}} \log \frac{1}{\varepsilon})$.
\end{theorem}

The randomization in the algorithm appears inside calls to Laplacian solvers, and can be removed at a multiplicative cost of $n^{o(1)}$ in running time~\cite{ChuzhoyDeterministic}.

Table~\ref{tab::comparison} compares the running times of algorithms for this problem.
For square matrices, ours becomes faster than the application of cutting planes~\cite{jiang2020improved}
  when $s < n^2$.
Our algorithm becomes faster than that of Objois and Vladu \cite{objois2026approximating} when $\varepsilon < 1/\sqrt{s}$.
For $s \leq \Otilde(n)$ and $\varepsilon \leq 1/n^2$, our algorithm is faster than both of these by a factor of $n^{3/2}$.

\begin{table}[ht]
\centering
\caption{Comparison of different algorithms for approximating the $p \to q$ norm of a non-negative matrix. Assume the matrix is square or moderately tall, i.e. $m \le O(n)$.}
\label{tab::comparison}
\begin{tabular}{lccccc}
\toprule
Algorithm & Runtime & $s \le \widetilde{O}(n)$ & $\varepsilon \le O(n^{-c})$ & $c = \frac{1}{2}$ & $c = 2$ \\
          &         & (sparse input)           & (high accuracy)             &                  &         \\
\midrule
\cite{bhaskara2011approximating} & $\widetilde{O}(\frac{n^3 s}{\varepsilon})$ & $\widetilde{O}(\frac{n^4}{\varepsilon})$ & $\widetilde{O}(n^{4+c})$ & $\widetilde{O}(n^\frac{9}{2})$ & $\widetilde{O}(n^{6})$ \\
\cite{objois2026approximating} & $\widetilde{O}(\frac{s}{\varepsilon})$ & $\widetilde{O}(\frac{n}{\varepsilon})$ & $\widetilde{O}(n^{1+c})$ & $\widetilde{O}(n^\frac{3}{2})$ & $\widetilde{O}(n^{3})$ \\
\cite{jiang2020improved} & $\widetilde{O}((ns + n^3) \log \frac{1}{\varepsilon})$ & $\widetilde{O}( n^3\log \frac{1}{\varepsilon})$ & $\widetilde{O}(n^3)$ & $\widetilde{O}(n^3)$ & $\widetilde{O}(n^{3})$ \\
This work & $\widetilde{O}(s^{\frac{3}{2}} \log \frac{1}{\varepsilon})$ & $\widetilde{O}(n^{\frac{3}{2}} \log \frac{1}{\varepsilon})$ & $\widetilde{O}(n^{\frac{3}{2}})$ & $\widetilde{O}(n^{\frac{3}{2}})$ & $\widetilde{O}(n^{\frac{3}{2}})$ \\
\bottomrule
\end{tabular}
\end{table}

\subsection{Techniques and proof overview}

We begin by formulating $\|\AA\|_{p\to q}$ as a linear optimization problem over a convex set that can be expressed as the intersection of affine constraints with a product of power cones. To the best of our knowledge, this is different from the convex formulations previously studied for this problem \cite{steinberg2005computation, objois2026approximating}.

This convex set has an efficient self concordant barrier function with complexity parameter bounded by $O(s)$ (Section \ref{sec::conic}). Therefore, we use a standard short-step path-following interior point method (IPM) to solve the resulting problem \cite{nesterov1994interior, renegar2001mathematical}, which converges to an $\varepsilon$-approximate solution in $O(\sqrt{s}\log s / \varepsilon)$ iterations.

As the convex set over which we optimize only depends on the non-zero structure of the input matrix $\AA$ but not on its entries, we can derive a closed form for the analytic center of the convex body over which we optimize.
This makes the initialization of the IPM straightforward (Lemmas \ref{lem::analytic-center} and \ref{lem::eta0}).
Termination is even easier because no rounding is required (Lemma \ref{lem::termination}).

The reason this leads to a fast algorithm is that we can use Laplacian solvers to quickly solve the linear equations that arise at each step of the IPM. More precisely, after eliminating a diagonal block in the Hessian matrix of the barrier function, the Schur complement can be decomposed as an M-matrix with known diagonal scaling, plus a rank-$2$ matrix (Theorem \ref{thm::hessian-schur}), which is exactly the type of linear systems that can be solved quickly by combining Laplacian solvers and Woodbury's matrix identity. To conclude the argument, it suffices to show that the error introduced by Laplacian solvers is small enough to produce a valid approximate Newton step, but large enough to solve the system quickly, for all iterations (Lemma \ref{lem::newtonsolve-works}).

The use of Laplacian solvers inside IPMs and the analysis of the impact of their approximate solves is now mature (See, e.g. \cite{daitch2008faster,madry2013navigating,lee2014path,kyng2015lipschitz,kyng2015fast,cohen2017negative,vandenbrand2020bipartite,vandenbrand2021minimum,gao2022fully,dong2022nested} and following work).
However, it is surprising that they can be applied here.
The previous applications of Laplacian solvers inside IPMs of which we are aware are for problems that have graph structure, and which concerned differences of functions across edges in a graph.
Neither edges nor differences of functions appear in our problem.

On top of the strong theoretical guarantees of Theorem \ref{thm::main}, our algorithm is also practical. In Section \ref{sec::experiments}, we discuss an implementation of Algorithm \ref{alg::IPM} which performs faster than commercial solvers such as MOSEK \cite{aps2024mosek} on a variety of matrices (both sparse and dense) and accuracy targets.

\subsection{Related work}

Aside from singular values, there are several other linear algebraic notions that have been generalized to $\ell_p$ geometries and studied under a computational lens to improve results based on numerical linear algebra primitives \cite{ovsiankin2026computation}. Some notable examples include $p$-norm regression \cite{bubeck2018homotopy, jambulapati2022improved, adil2024fast}, and $p$-Laplacian eigenvalues \cite{amghibech2003eigenvalues, buhler2009spectral, hein2010inverse}. The $p$-norm regression problem can be thought of as a generalization of the least squares problem when $p=2$, while the $p$-Laplacian eigenvalues correspond to the second smallest Laplacian eigenvalue when $p=2$. Fast algorithms have been designed for the special case of $p$-norm regression restricted to graphs ($p$-norm flows), see e.g. \cite{kyng2015fast, adil2020faster, chen2025maximum}.

\medskip

The first published polynomial-time algorithm for the $p\to q$ matrix norm problem was given by Bhaskara and Vijayaraghavan \cite{bhaskara2011approximating}, and it consists of a non-linear power method originally proposed by Boyd \cite{boyd1974power}. The nearly-linear time algorithm by Objois and Vladu \cite{objois2026approximating} is inspired by the non-linear power method but it implements a much more efficient potential function argument. To the best of our knowledge, all current high-accuracy solvers for this problem are based on applying techniques for general convex optimization to the convex formulation from \cite{steinberg2005computation}, and our work is the first to explicitly leverage the structure of the feasible convex set.

%% file: sections/02_preliminaries.tex
\section{Preliminaries}

\subsection{Notation and linear algebra background}
We will use boldface capital letters to denote matrices, e.g. $\AA \in \mathbb{R}^{m \times n}$ and boldface lower case letters to denote vectors, e.g. $\zz \in \mathbb{R}^m$.  We denote the inner product of vectors $\xx,\yy \in \mathbb{R}^n$ by $\langle \xx,\yy\rangle = \sum_{i=1}^n \xx(i)\yy(i)$. We let  $\Delta_n$ denote the $n$-dimensional simplex, $\{\xx \in \mathbb{R}^n: \xx \ge 0, \sum_{i=1}^n \xx(i) \le 1\}$. A square symmetric matrix $\AA \in \mathbb{R}^{n \times n}$ is positive definite, written $\AA \succ 0$, if all its eigenvalues are positive. Given a positive definite matrix $\AA$, the $\AA$-norm is defined as 
\begin{equation*}
    \|\xx\|_{\AA} = \sqrt{\xx^\top \AA \xx} = \|\AA^\frac{1}{2}\xx\|_2
\end{equation*}

We will use big-$O$ notation throughout to emphasize asymptotics in upper bounds. The notation $\widetilde
O$ hides polynomials in logarithms of the size parameters ($m$, $n$, $s$), but not the accuracy parameters ($\delta$, $\varepsilon$).

\subsubsection{Dual norms}

For $p > 1$,  the $p$-norm of a vector $\xx$ is given by 
\[
\norm{\xx}_p \defeq \left( \sum_i \abs{\xx(i)}^p \right)^{1/p}.
\]
Throughout this paper we let $p^{*}$ be the solution to the equation 
\[
\frac{1}{p} + \frac{1}{p^{*}} = 1.
\]
The dual of the $p$-norm is the $p^{*}$-norm.
It satisfies
\[
\norm{\xx}_p = \max_{\norm{\yy}_{p^{*}} = 1} \langle \xx, \yy \rangle.
\]

\subsubsection{SDD matrices, M-matrices, and Laplacian solvers}

A PSD matrix $\MM$ is an M-matrix if it has non-positive off-diagonal entries. A symmetric matrix $\AA$ is diagonally dominant (SDD) if for all $j$, 
  $\AA(j,j) \geq \sum_{i \neq j} \abs{\AA(i,j)}$.
If a positive definite matrix $\MM \succ 0$ is an M-matrix, then there exists a diagonal matrix $\DD$ with positive diagonal entries such that $\DD \MM \DD$ is SDD. 

Systems of linear equations in SDD matrices can be solved in nearly-linear time~\cite{spielman2004nearly,koutis2014approaching,cohen2014solving,jambulapati2025ultrasparse}.
We make critical use of the fact that the algorithms for solving these are in fact linear operators.

The following theorem summarizes these results in a form similar to that which appears in Kyng et al. \cite[Theorem A.4]{kyng2015fast}.
\begin{theorem}\label{thm::laplacian-solvers}
    Let $\LL \in \mathbb{R}^{n \times n}$ be a positive definite SDD matrix with $s$ non-zero entries and $0 < \delta < 1$.
There exists an algorithm \textsc{STSolve} that takes $\LL$ and $\delta$ as input and in expected time $\widetilde{O}(s)$ computes a program \textsc{solveL},
    such that for every vector $\bb \in \mathbb{R}^n$
\[
  \textsc{solveL}(\bb) = \PP \bb,
\]
for a symmetric linear operator $\PP$ such that 
    \begin{equation*}
        (1 - \delta) \LL^{-1} \preceq \PP \preceq (1+\delta)\LL^{-1}.
    \end{equation*}
The program \textsc{solveL} runs 
    in time $\widetilde{O}\left(s \log\frac{1}{\delta}\right)$. 
\end{theorem}

Deterministic variations are available at the cost of an additional factor of $n^{o(1)}$ in running time~\cite{ChuzhoyDeterministic}.

\subsubsection{Woodbury identity}

We will encounter matrices that can be expressed as the sum of SDD matrices with matrices of rank 2. It states that for an invertible matrix $\AA \in \mathbb{R}^{n \times n}$ and matrices $\UU \in \mathbb{R}^{n \times k}, \CC \in \mathbb{R}^{k \times k}, \VV \in \mathbb{R}^{k \times n}$ with $\CC$ invertible, the matrix $\AA + \UU \CC \VV$ is invertible if and only if $\CC^{-1} + \VV \AA^{-1}\UU$ is invertible, and it satisfies
\begin{equation*}
    (\AA + \UU \CC \VV)^{-1} = \AA^{-1} - \AA^{-1}\UU(\CC^{-1} + \VV \AA^{-1}\UU)^{-1} \VV \AA^{-1}
\end{equation*}
In particular, when $\CC = I$ and $\VV = \UU^\top$, we have
\begin{equation}\label{eq::woodbury}
    (\AA + \UU \UU^\top)^{-1} = \AA^{-1} - \AA^{-1}\UU(I + \UU^\top \AA^{-1}\UU)^{-1} \UU^\top \AA^{-1}
\end{equation}

\subsubsection{Schur complement}
Suppose we want to solve equations in a positive definite  square matrix $\AA \in \mathbb{R}^{(n_1 + n_2) \times (n_1 + n_2)}$ 
\begin{align*}
    \AA = \begin{pmatrix}
        \AA(S_1,S_1) & \AA(S_1,S_2) \\
        \AA(S_2,S_1) & \AA(S_2,S_2)
    \end{pmatrix}
\end{align*}
where the rows and columns are partitioned in two blocks $S_1, S_2$ such that $|S_1| = n_1$ and $|S_2| = n_2$. If the matrix $\AA(S_2,S_2)$ is invertible, then the Schur complement of $\AA$ with respect to $S_2$ is defined as
\begin{equation}\label{eq::schur-definition}
    \AA \setminus S_2 = \AA(S_1,S_1) - \AA(S_1,S_2) \AA(S_2,S_2)^{-1} \AA(S_2,S_1)
\end{equation}

The problem of solving systems of linear equations in $\AA$ can be reduced to that of solving a system of equations in $\AA \setminus S_2$ followed by a system of equations in $\AA(S_2, S_2)$:

\begin{align*}
    \begin{pmatrix}
        \AA(S_1,S_1) & \AA(S_1,S_2) \\
        \AA(S_2,S_1) & \AA(S_2,S_2)
    \end{pmatrix}
    \begin{pmatrix}
        \xx_1 \\
        \xx_2
    \end{pmatrix}
    = 
    \begin{pmatrix}
        \bb_1 \\
        \bb_2
    \end{pmatrix}
    \iff
    (\AA \setminus S_2) \xx_1 &= \bb_1 - \AA(S_1,S_2)\AA(S_2,S_2)^{-1} \bb_2 \\
    \xx_2 &= \AA(S_2, S_2)^{-1} (\bb_2 - \AA(S_2,S_1) \xx_1)
\end{align*}

This decomposition provides efficiency gains when systems in $\AA(S_2, S_2)$ are easy to solve, such as when it is diagonal, and when systems in the Schur complement $\AA \setminus {S_2}$ are easy to solve.

We will use the following variational characterization of the Schur complement from \cite{boyd2004convex}.
\begin{lemma}\label{lem::variational-schur}
    For any $\xx_1 \in \mathbb{R}^{S_1}$
    \begin{equation*}
        \xx_1^\top (\AA \setminus S_2) \xx_1 = 
        \min_{\xx_2 \in \mathbb{R}^{S_2}} 
        \begin{pmatrix}
            \xx_1 \\
            \xx_2
        \end{pmatrix}^\top
        \AA
        \begin{pmatrix}
            \xx_1 \\
            \xx_2
        \end{pmatrix}
    \end{equation*}
    and the minimum is attained at $\xx_2 = - \AA(S_2,S_2)^{-1}\AA(S_2,S_1) \xx_1$
\end{lemma}

\subsection{Convex optimization}\label{sec::background_IPM}

\subsubsection{Self-concordant barrier functions}

For a smooth function $F:\mathbb{R}^n \to \mathbb{R}$ with Hessian $H(x) = \nabla^2 F(x)$, we define the Hessian norm of a vector $\hh \in \mathbb{R}^n$ as $\|\hh\|_{\xx} = \sqrt{\hh^\top \HH(\xx) \hh}$.
\begin{definition}[\cite{nesterov1994interior}]
    Let $\mathcal{K} \subseteq \mathbb{R}^n$ be a closed and convex set with nonempty interior. A smooth function $F: int(\mathcal{K}) \to \mathbb{R}$ is \textit{self-concordant} if its third derivative is uniformly bounded by a twice the Hessian norm cubed.
    \begin{align*}
        |D^3F(x)[h,h,h] | \le 2\|h\|_{\nabla^2 F(x)}^3 \qquad \forall x \in int(\mathcal{K}), h \in \mathbb{R}^n
    \end{align*}
    Moreover, $F$ is a $\nu$-self-concordant \textit{barrier} function if it diverges to $+\infty$ at the boundary and 
    \begin{align*}
        |\langle \nabla F(x), h\rangle| \le \sqrt{\nu} \|h\|_{\nabla^2 F(x)}
    \end{align*}
\end{definition}
The constant $\nu>0$ is often called the ``complexity parameter'' of $F$. For any $\xx \notin int(\mathcal{K})$, we extend the barrier function as $F(\xx) = +\infty$. We list some useful properties of self-concordant barrier functions that we will use throughout this paper, see \cite[Proposition 2.3.1]{nesterov1994interior}.
\begin{enumerate}
    \item[(\textbf{SC1})] For an affine map $T(x) = Ax + b$, the function $F_1(x) = F(T(x))$ is $\nu$-self-concordant \label{itm::sc1}
    \item[(\textbf{SC2})] For two self-concordant functions $F_1(x)$ and $F_2(x)$ with complexity parameters $\nu_1$ and $\nu_2$, respectively, the function $F(x) = F_1(x) + F_2(x)$ is $(\nu_1 + \nu_2)$-self-concordant. \label{itm::sc2}
\end{enumerate}

\subsubsection{Interior Point Methods}

Interior point methods (IPMs) were introduced by the seminal works of Dikin \cite{dikin1967iterative} and Karmarkar \cite{karmarkar1984new} to solve linear programming problems efficiently, and were later extended by Nesterov and Nemirovskii \cite{nesterov1994interior} to general convex optimization. Let $\mathcal{K} \subseteq \mathbb{R}^n$ be a closed convex set and consider the constrained linear optimization problem, for a fixed $\cc \in \mathbb{R}^n$,
\begin{equation*}
    \xx^* = \arg \min_{\xx \in \mathcal{K}} \langle \cc, \xx \rangle
\end{equation*}
Let $\eta > 0$ and $F:\mathcal{K} \to \mathbb{R}$ be a self-concordant barrier function for $\mathcal{K}$ with complexity parameter $\nu$, and consider the regularized (unconstrained) linear problem $\Phi_\eta(\xx)$ with \textit{central path} $\xx_\eta$
\begin{align*}
    \Phi_\eta(\xx) &\defeq F(\xx) + \eta \langle \cc, \xx \rangle \\
    \xx_\eta^* &\defeq \arg \min_{\xx \in \mathbb{R}^n} \Phi_\eta(\xx)
\end{align*}

The intuition behind the IPM approach is to use the barrier function to ensure the problem stays feasible, while smoothly penalizing points that are close to the boundary of $\mathcal{K}$. When the central path parameter $\eta$ is small, $\Phi_\eta$ ``looks like'' $F$, so the central path will be away from the boundary. On the other hand, when $\eta$ is large, $\Phi_\eta$ encodes more structure from the linear component $\langle \cc, \xx \rangle$, so the central path will be closer to the boundary in the proximity of the (constrained) optimum. If the function $F$ is easy to minimize, then a natural strategy for approaching $x^*$ (and the one used successfully by IPMs) is to start close to the minimizer of $F$ (also called the \textit{analytic center} of $\mathcal{K}$), and then follow the central path while increasing $\eta$.

The iterative phase of interior point method follows the central path $\{\xx_\eta^*\}_{\eta \ge 0}$ by alternating Newton updates for $\xx$ and multiplicative updates for the central path parameter $\eta$. Let $\HH(\xx) = \nabla^2 F(\xx)$ and recall Newton's method for function minimization,
\footnote{Don't confuse the simplex $\Delta_n$ with the  \textit{Newton step} $\Delta_{\eta}(\xx) = - \HH(\xx)^{-1} \nabla \Phi_\eta(\xx)$.}
\begin{equation*}
    \xx_+ = \xx - \HH(\xx)^{-1} \nabla \Phi_\eta(\xx) \defeq \xx + \Delta_\eta (\xx)
\end{equation*}
Define the \textit{Newton decrement} as the Hessian norm of the Newton step
\begin{equation*}
    \lambda_\eta(\xx) = \|\HH(\xx)^{-1} \nabla \Phi_\eta (\xx)\|_{\xx} = \|\Delta_\eta(\xx)\|_{\xx}
\end{equation*}
This controls the quadratic convergence region of Newton's method, which is the key invariant used to analyze the convergence of IPMs. We will be interested in the convergence of Newton's method under approximate Newton steps $\widetilde{\Delta}_\eta(\xx)$ in the following sense:
\begin{equation}\label{eq::newton-approx}
    \|\widetilde{\Delta}_\eta(\xx) - \Delta_\eta(\xx)\|_{\xx} \le \delta \|\Delta_\eta(\xx)\|_{\xx}
\end{equation}

There are many different versions of IPMs, and the following theorem suffices for our application. See Appendix \ref{sec::app_IPM} for more details.
\begin{theorem}\label{thm::IPM_full}
    Let $\xx_0 \in int(\mathcal{K})$ and $0 < \eta_0 < 1$ satisfy $\lambda_{\eta_0}(\xx_0) \le \frac{1}{10}$, $0 < \varepsilon < 1$, and $0 < \delta < \frac{1}{5}$. Then there exists a constant $C > 1$ such that after $\left\lceil C \sqrt{\nu} \log\frac{\nu}{\varepsilon \eta_0} \right\rceil$ iterations each involving a $\delta$-approximate Newton step in the sense of \eqref{eq::newton-approx}, we reach a point $\xx \in int(\mathcal{K})$ satisfying
    \begin{equation*}
        \langle \cc, \xx\rangle - \langle \cc, \xx^*\rangle \le \varepsilon
    \end{equation*}
    Moreover, if $\nu \ge 40$ we can set $C = 22$.
\end{theorem}

The typical choice for $\xx_0$ is an approximation of $\xx_0^*$, which is also called the \textit{analytic center}. This is convenient because it is the global minimizer of the self-concordant barrier $F$, and it does not depend on $\eta$. The initial $\eta_0$ is then chosen to be as large as possible while ensuring $\lambda_{\eta_0}(\xx_0) \le \frac{1}{10}$. To turn Theorem \ref{thm::IPM_full} into an efficient algorithm, we need $F$ to be an \textit{efficient} self-concordant barrier for $\mathcal{K}$, i.e. we need to be able to compute gradient and Hessian efficiently.

\subsubsection{The power cone}
For $0 < \alpha < 1$, the $3$-dimensional power cone is defined as 
\begin{equation*}
    \mathcal{K}_{\alpha} = \{\xx \in \mathbb{R}^3: \xx(1)\ge 0, \xx(2)\ge 0, |\xx(3)| \le \xx(1)^\alpha \xx(2)^{(1-\alpha)}\}
\end{equation*}
The following function is a self-concordant barrier for the power cone with complexity parameter $\nu = 3$, which is a result of Chares \cite{chares2009cones}. 
\begin{equation*}
    F_\alpha(\xx) = -\log\left(\xx(1)^{2\alpha} \xx(2)^{2(1-\alpha)} - \xx(3)^2\right) - (1-\alpha)\log(\xx(1)) - \alpha\log(\xx(2))
\end{equation*}

For positive $\alpha,\beta$ with $\alpha+\beta < 1$, the $4$-dimensional generalized power cone is similarly defined.
\begin{equation*}
    \mathcal{K}_{\alpha, \beta} = \{\xx \in \mathbb{R}^4: \xx(1)\ge 0, \xx(2)\ge 0, \xx(3) \ge 0, |\xx(4)| \le \xx(1)^\alpha \xx(2)^\beta \xx(3)^{(1-\alpha-\beta)}\}
\end{equation*}
Roy and Xiao \cite{roy2022self} proved that the following generalization of Chares' function is self-concordant for $\mathcal{K}_{\alpha,\beta}$ with complexity parameter $\nu=4$.\footnote{They prove a more general statement, but this suffices for our applications.}
\begin{align}\label{eq::scf-power}
    F_{\alpha,\beta}(\xx) = &-\log\left(\xx(1)^{2\alpha} \xx(2)^{2\beta} \xx(3)^{2(1-\alpha-\beta)} - \xx(4)^2\right) \\
    &- (1-\alpha)\log(\xx(1)) - (1-\beta)\log(\xx(2)) - (\alpha+\beta)\log(\xx(3)) \nonumber
\end{align}

%% file: sections/03_conic.tex
\section{A convex formulation of non-negative matrix norm}\label{sec::conic}

Let $p \ge q > 1$ and let $\AA \in \mathbb{R}_{\ge 0}^{m \times n}$ be a non-negative matrix. 
We formulate the problem of computing $\|\AA\|_{p \to q}$ as a linear optimization problem over a convex set.
Following \cite[Example 3.1]{boyd2004convex}, we write 
\begin{equation*}
       \|\AA\|_{p \to q} = \max_{\uu \in \mathbb{R}^n:\|\uu\|_p=1} \|\AA \uu\|_q = \max_{\substack{\uu \in \mathbb{R}^n:\|\uu\|_p=1\\ v \in \mathbb{R}^m: \|\vv\|_{q^*}=1}} \vv^\top \AA \uu ,
\end{equation*}
which follows from the definition of dual norms. 
Since the matrix $\AA$ is non-negative, $\vv^\top \AA \uu \le |\vv|^\top \AA |\uu|$, so the maximizers are non-negative.
We then make the entrywise change of variables $\yy=\uu^p, \zz = \vv^{q^*}$ 
to obtain the form
\begin{align*}
    \|\AA\|_{p \to q} &= \max_{\substack{\uu \in \mathbb{R}^n:\|\uu\|_p \le 1\\ \vv \in \mathbb{R}^m: \|\vv\|_{q^*} \le 1}} \sum_{i=1}^m \sum_{j=1}^n \AA(i,j) \vv(i) \uu(j)
    =& \max_{\substack{\yy \in \Delta_n\\ \zz \in \Delta_m}} \sum_{i=1}^m \sum_{j=1}^n \AA(i,j) \zz(i)^\frac{1}{q^*} \yy(j)^{\frac{1}{p}} .
\end{align*}

We now linearize the objective function while preserving the program's value by introducing additional variables $\GGamma$:
\begin{align}\label{eq::norm-conic}
    \|\AA\|_{p \to q} =& \max_{\zz \ge 0, \yy \ge 0, \GGamma} \sum_{i=1}^m \sum_{j=1}^n \AA(i,j) \GGamma(i,j) \\
    \text{s.t. }& \sum_{i=1}^m \zz(i) \le 1 \nonumber \\
    & \sum_{j=1}^n \yy(j) \le 1 \nonumber \\
    & |\GGamma(i,j)| \le \zz(i)^\frac{1}{q^*} \yy(j)^\frac{1}{p} & \forall i=1,...,m, j=1,...,n \nonumber
\end{align}

We can focus on the non-zero entries of $\AA$, since if $\AA(i,j)=0$ the contribution of $\AA(i,j)\Gamma(i,j)$ is $0$ and is achieved by $\GGamma(i,j)=0$. If $\AA$ has rows or columns that are identically zero, then we can remove them without changing either the norm or the conic program, since the corresponding $\zz(i)$ or $\yy(j)$ can be set to zero. 
Let $S$ be the set of non-zero entries of $\AA$, with $s = |S|$, and write $i \sim j, j \sim i$ if $(i,j) \in S$. Then, the program above has $m+n+s$ variables and $s$ power cone constraints and $2$ affine constraints, 
where $\max(m,n) \le s \le nm$. In the rest of the paper, we always assume we are working with sparse representations of matrices. For $(i,j)$ such that $\AA(i,j) > 0$, let $k_{i,j} \in \{1,...,s\}$ be the index of the pair $(i,j)$ when ordering the non-zero entries of $\AA$ lexicographically (i.e. flattening row-wise). 

Let $\mathcal{K}$ be the feasible set of the optimization problem \eqref{eq::norm-conic}, then the next lemma shows that $\mathcal{K}$ is convex and has an efficient self-concordant barrier.
\begin{lemma}
    For any $p \ge q > 1$ and non-negative matrix $\AA\in \mathbb{R}^{m \times n}$, the set $\mathcal{K} \subseteq \mathbb{R}^{m+n+s}$ defined in \eqref{eq::norm-conic} is convex and the function $F:int(\mathcal{K}) \to \mathbb{R}$
    \begin{align}\label{eq::scf_barrier_F}
        F(\zz,\yy,\ggamma) = - \sum_{(i,j) \in S} \left( \log \left( \zz(i)^\frac{2}{q^*} \yy(j)^\frac{2}{p} - \ggamma(i,j)^2\right) + \frac{1}{q}\log \zz(i) + \frac{1}{p^*} \log \yy(j)\right)  \\
        - \log(1-\sum_{i'}\zz(i')) - \log(1-\sum_{j'}\yy(j')) \nonumber
    \end{align}
    is a self-concordant barrier function for $\mathcal{K}$ with complexity parameter $\nu = 3s+2$.
\end{lemma}
\begin{proof}
    Let $\alpha = \frac{1}{q^*}$ and $\beta = \frac{1}{p}$. The constraint on $\GGamma(i,j)$, $\zz(i)$, and $\yy(j)$, together with non-negativity of $\zz(i)$ and $\yy(j)$, correspond to membership in the following affine slice of the $4$-dimensional power cone
    \begin{equation}\label{eq::power-single}
        (\zz(i), \yy(j), 1, \GGamma(i,j)) \in \mathcal{K}_{\alpha,\beta}
    \end{equation}
    This is equivalent to adding an extra variable and just setting it to one. Consider the affine map $P_{i,j}(\xx) = (\xx(i), \xx(m+j), 1, \xx(m+n+k_{i,j}))$, then the set \eqref{eq::norm-conic} has the following conic (hence convex) 
    
    \begin{equation}\label{eq::feasible-set}
        \mathcal{K} = \left\{\xx \in \mathbb{R}^{m+n+s}: \forall (i,j)\in S, P_{i,j}(\xx) \in \mathcal{K}_{\alpha,\beta}, \sum_{i=1}^m \xx(i) \le 1, \sum_{j=1}^n \xx(m+j) \le 1\right\}
    \end{equation}
    
    The set \eqref{eq::feasible-set} has a natural (efficient) self-concordant barrier function which is obtained by summing the barrier functions for each individual power cone, i.e. $F_{\alpha,\beta}$ \eqref{eq::scf-power} for $\mathcal{K}_{\alpha,\beta}$ and the log barrier for the linear inequalities, which is exactly \eqref{eq::scf_barrier_F}.\footnote{When $p=q$, technically we use the self-concordant barrier of Chares since $\mathcal{K}_{\alpha,\beta}$ is defined for $p > q$. However, the rest of the argument and the analysis are unchanged.} 
    Since \eqref{eq::power-single} is just an affine restriction of the power cone, its complexity parameter is at most $4$, i.e. the complexity of the barrier on the $4$-dimensional space (\hyperref[itm::sc1]{SC1}). We can improve this bound to $3$.
    
    Denoting $\tt = (z,y,1,\gamma)$, $G=F_{\alpha,\beta}$, $\HH = \nabla^2 G(\tt)$, $U=\{1,2,4\}$, and $\BB = \HH(U,U)$. Since $G(c\tt) = G(\tt) - 4\log c$, we have $\nabla G(\tt) = -\HH \tt$ and $\tt^\top \HH \tt = 4$. Moreover, for any vector $\hh\in \mathbb{R}^4$ with $\hh(3)=1$, $\tt - \hh \notin int(\mathcal{K})$, so the Dikin ellipsoid property \cite[Theorem 2.1.1]{nesterov1994interior} implies $\hh^\top \HH \hh \ge 1$. By Lemma \ref{lem::variational-schur}, this is is a lower bound to the the scalar Schur complement $\sigma = \HH \setminus U$. Define $f(x,y,\gamma) = G(\tt)$, then completing teh square gives
    \begin{align*}
        \nabla f^\top \BB^{-1} \nabla f = \tt^\top \HH \tt - \sigma = 4 - \sigma \le 3
    \end{align*}
    This means that each restricted barrier has complexity parameter at most $3$, and summing over all barriers and using (\hyperref[itm::sc2]{SC2}) shows that the complexity parameter of $F$ satisfies $\nu \le 3s+2$.
\end{proof}

Consider the vectors $\aa = vec(\AA) \in \mathbb{R}^s$, and $\ggamma = vec(\GGamma) \in \mathbb{R}^s$. Since they are vector representations of matrices, we will abuse notation and adopt two-dimensional indexing to make the edge structure explicit, e.g. $\ggamma(i,j)=\ggamma(k_{i,j})$. Denoting $\Bar{\aa} = (0,0,\aa) \in \mathbb{R}^{m+n+s}$, we can write the convex optimization problem \eqref{eq::norm-conic} as
\begin{align*}
    \max &\langle \bar{\aa}, \xx\rangle \hspace{0.5cm} = \hspace{0.4cm}  -\min\langle-\bar{\aa},\xx\rangle \\
    \text{s.t. }& \xx \in \mathcal{K} \hspace{1.3cm} \text{s.t. } \xx \in \mathcal{K}
\end{align*}

%% file: sections/04_IPMsec.tex
\section{An interior point method for \texorpdfstring{$\|A\|_{p \to q}$}{}}\label{sec::IPM}

Interior point methods (IPM) are a family of algorithms designed to solve linear (or more generally, convex) optimization problems over convex sets with efficient self-concordant barriers (see Section \ref{sec::background_IPM}). Since \eqref{eq::norm-conic} is a conic program with an efficient barrier, it is natural to consider an algorithm based on IPM for it, which is described in Algorithm \ref{alg::IPM}. See Appendix \ref{sec::app_derivatives} for the closed form of the derivatives of the self-concordant barrier \eqref{eq::scf_barrier_F}.

\begin{algorithm}[H]
\SetAlgoLined
\caption{\textsc{pqNormIPM}}\label{alg::IPM}
\SetKwInOut{Input}{Input}
\SetKwInOut{Output}{Output}
\Input{Non-negative matrix $\AA \in \mathbb{R}^{m \times n}$, accuracy parameter $\varepsilon > 0$, exponents $p \ge q > 1$.}
\Output{Vector $\uu \in \mathbb{R}^{n}$ such that $\|\uu\|_p=1$ and $\|\AA \uu\|_q \ge (1-\varepsilon)\|\AA\|_{p\to q}$}

If $\AA=0$, return $\uu = (n^{-\frac{1}{p}})\mathbf{1}_n$
\\
Pre-process the matrix $\AA$ by removing all-zero rows and columns, and let $s$ be the number of non-zero entries left in $\AA$.
\\
Compute $A_{\max} = \max_{(i,j)\in S}\AA(i,j)$ and set $\AA' = \frac{1}{A_{\max}}A$. 
\\
Set $\delta = \frac{1}{6}$ and $T = \lceil 100 \sqrt{s} \log\frac{s}{\varepsilon} \rceil$ 
\\
Compute $d_i = |\{j:(i,j)\in S\}|$ for each $i \in \{1,...m\}$, and $e_j = |\{i:(i,j)\in S\}|$ for each $j \in \{1,...n\}$ 
\\
For each $i \in \{1,...m\}$, $j \in \{1,...n\}$, and $(i,j) \in S$, initialize
\begin{align*}
    \zz_0(i) &= \frac{\left(1 + \frac{1}{q^*} \right) d_i}{1 + \left(1 + \frac{1}{q^*} \right)s}
    \qquad
    \yy_0(j) &= \frac{\left(1 + \frac{1}{p} \right) e_j}{1 + \left(1 + \frac{1}{p} \right)s} 
    \qquad
    \ggamma_0(i,j) &= 0 
    \qquad    
    \eta_0 &= \frac{\sqrt{2}}{10\sqrt{s}} 
\end{align*}
\\
\For{$t = 0,...,T-1$} {
$(\zz_{t+1}, \yy_{t+1}, \ggamma_{t+1}) = (\zz_{t}, \yy_{t}, \ggamma_{t}) + \textsc{NewtonSolve}(\AA', \delta, (\zz_{t}, \yy_{t}, \ggamma_{t}), \eta_t)$  (Algorithm \ref{alg::NewtonSolve})
\\
$\eta_{t+1} = \left( 1 + \frac{1}{20 \sqrt{3s+2}}\right) \eta_t$
}
$Y_T = \sum_{j=1}^n \yy_T(j)$
\\
\Return{$\uu = \left(\frac{1}{Y_T}\yy_T\right)^{\frac{1}{p}} $}
\end{algorithm}

\subsection{Initialization}

Without loss of generality, we assume $\max_{(i,j) \in S} \AA(i,j) \le 1$. Otherwise we can first compute $A_{\max} = \max_{(i,j) \in S} \AA(i,j)$ in linear time and work with the matrix $\AA' = \frac{1}{A_{\max}} \AA$ (Step 3 of Algorithm \ref{alg::IPM}).

The analytic center for \eqref{eq::scf_barrier_F} can be computed in closed form in linear time. Let $d_i$ be the number of non-zero entries in the $i$-th row of $A$ and $e_j$ be the number of non-zeros in the $j$-th column. Note that $\sum_{i=1}^m d_i = \sum_{j=1}^n e_j = s$.
\begin{lemma}\label{lem::analytic-center}
    The minimizer for the self-concordant barrier \eqref{eq::scf_barrier_F} is given by $\xx_0 = (\zz_0, \yy_0, \ggamma_0)$, where
    \begin{align*}
        \zz_0(i) = \frac{\left(1 + \frac{1}{q^*} \right) d_i}{1 + \left(1 + \frac{1}{q^*} \right)s} \qquad
        \yy_0(j) = \frac{\left(1 + \frac{1}{p} \right) e_j}{1 + \left(1 + \frac{1}{p} \right)s} \qquad
        \ggamma_0(i,j) = 0
    \end{align*}
    This can be computed in time $O(s)$.
\end{lemma}
\begin{proof}
    Since the function $F$ is strictly convex and twice differentiable, its minimizer is unique and can be found by solving the first-order conditions $\nabla F(\zz_0,\yy_0,\ggamma_0) = 0$. Setting the gradient with respect to $\gamma$  to zero forces $\ggamma_0(i,j)=0$ for each $(i,j) \in S$ (see equation \eqref{eq::grad_gam}). This simplifies the remaining derivatives, for example the gradient with respect to $\zz$  becomes (see equation \eqref{eq::grad_z})
    \begin{equation*}
        \left.\frac{\partial F}{\partial \zz(i)}\right|_{\gamma=0} = - \left(1 + \frac{1}{q^*} \right)\frac{d_i}{\zz(i)} + \frac{1}{1-Z} 
    \end{equation*}
    where $Z = \sum_{i=1}^m z(i)$. Crucially, this does not depend on $\yy$, and we can solve it by isolating $\zz(i)$ and summing the equations to deal with the $Z$ term.
    \begin{align*}
        \zz(i) = \left(1 + \frac{1}{q^*} \right)(1-Z) d_i& \iff Z = \left(1 + \frac{1}{q^*} \right)(1-Z) s \\
        &\iff Z = \frac{\left(1 + \frac{1}{q^*} \right) s}{1 + \left(1 + \frac{1}{q^*} \right)s} \\
        &\iff \zz(i) = \frac{\left(1 + \frac{1}{q^*} \right) d_i}{1 + \left(1 + \frac{1}{q^*} \right)s}
    \end{align*}
    The formula for $\yy(j)$ follows from identical calculations starting from \eqref{eq::grad_y}, and the fact that they can be computed in linear time follows simply by the fact that computing $d_i$ and $e_j$ for all $i,j$ requires a single scan over the matrix $A$.
\end{proof}

We can use this to initialize the parameter $\eta_0$ to order $s^{-\frac{1}{2}}$.
\begin{lemma}\label{lem::eta0}
    If we set $\xx_0$ as in Lemma \ref{lem::analytic-center} and choose
    \begin{equation*}
        \eta_0 = \frac{\sqrt{2}}{10\sqrt{s}},
    \end{equation*}
    then $\lambda_{\eta_0}(\xx_0) \le \frac{1}{10}$.
\end{lemma}
\begin{proof}
    We want to set $\eta_0$ such that $\lambda_{\eta_0}(\xx_0) = \frac{1}{10}$, for the $\xx_0=(\zz_0, \yy_0, \ggamma_0)$ defined in Lemma \ref{lem::analytic-center}. Since $\xx_0$ is the exact minimizer of $F(\xx)$, we have $\nabla F(\xx_0)=0$, hence $\nabla \Phi_\eta (\xx_0) = -\eta (0,0,\aa) = -\eta \Bar{\aa}$.
    \begin{align*}
        \lambda_\eta (\xx_0)^2 &= \nabla \Phi_\eta (\xx_0)^\top \HH(\xx_0)^{-1} \nabla \Phi_\eta (\xx_0) \\
        &= \eta^2 \Bar{\aa}^\top \HH(\xx_0)^{-1} \Bar{\aa}
    \end{align*}
    Note that at the analytic center, the block-decomposition of the Hessian \eqref{eq::hess_schur} has zero off-diagonal blocks, so the quadratic form above becomes
    \begin{align*}
        \Bar{\aa}^\top \HH(\xx_0)^{-1} \Bar{\aa} &=
        \begin{pmatrix}
            0 \\
            \aa
        \end{pmatrix}^\top
        \begin{pmatrix}
            \HH_w(\xx_0) & 0 \\
            0 & \HH_\gamma(\xx_0)
        \end{pmatrix}^{-1}
        \begin{pmatrix}
            0 \\
            \aa
        \end{pmatrix} \\
        &= \aa^\top \HH_\gamma(\xx_0)^{-1} \aa \\
        &= \frac{1}{2}\sum_{(i,j) \in S} \AA(i,j)^2 \zz(i)^\frac{2}{q^*} \yy(j)^\frac{2}{p}
    \end{align*}
    The last equality follows from the fact that $\HH_\gamma(\xx_0)$ is diagonal with entries given by \eqref{eq::hessian_gam2}. Using the fact that $\max_{(i,j) \in S} \AA(i,j) \le 1$,
    \begin{align*}
        \lambda_\eta (\xx_0)^2 &= \frac{\eta^2}{2} \sum_{(i,j) \in S} \AA(i,j)^2 \zz(i)^\frac{2}{q^*} \yy(j)^\frac{2}{p} \\
        &\le \frac{\eta^2}{2} \sum_{(i,j) \in S} \AA(i,j)^2 \le \frac{\eta^2 s}{2}
    \end{align*}
    To get $\lambda_{\eta_0}(\xx_0) \le \frac{1}{10}$, it suffices to choose 
    \begin{equation*}
        \eta_0 = \frac{\sqrt{2}}{10\sqrt{s}}
    \end{equation*}
    As desired.
\end{proof}

\subsection{Termination}
\begin{lemma}\label{lem::termination}
    The vector $\uu$ computed by Algorithm \ref{alg::IPM} satisfies
    \begin{equation*}
        \frac{\|\AA \uu\|_q}{\|\uu\|_p} \ge (1-\varepsilon)\|\AA\|_{p \to q}
    \end{equation*}
\end{lemma}
\begin{proof}
    Combining Theorem \ref{thm::IPM_full} and Lemma \ref{lem::eta0}, and using the fact that $\nu = 3s+2$ and $\delta=\frac{1}{6} < \frac{1}{5}$, we conclude that after $\left\lceil 22(\sqrt{3s+2}) \log\frac{10(3s+2)\sqrt{s}}{\sqrt{2}\varepsilon}\right\rceil \le \left\lceil 100 \sqrt{s} \log\frac{s}{\varepsilon} \right\rceil$ iterations of the IPM,\footnote{Whenever $s \ge 20$, this bound holds for any $0 < \varepsilon < 1$. We ignore issues of small $s$ since we can deal with it separately and solve it in constant time to inverse polynomial accuracy.} we find a feasible point $\xx=(\zz,\yy,\ggamma)$ satisfying 
    \begin{align*}
        &\sum_{(i,j) \in S} \AA'(i,j) \ggamma(i,j) \ge \|\AA'\|_{p \to q} - \varepsilon \\
        &\sum_{i=1}^m \zz(i) \le 1 \\
        &\sum_{j=1}^n \yy(j) \le 1 \\
        &\ggamma(i,j) \le \zz(i)^\frac{1}{q^*} \yy(j)^\frac{1}{p} & \forall (i,j) \in S
    \end{align*}
    where $\AA' = \frac{1}{A_{\max}}\AA$ has maximum entry at most one.\footnote{If $\ggamma$ has negative entries, we can safely change their sign since that can only improve the objective.} Recall that $A_{\max} \le \|\AA\|_{p \to q}$ for any non-negative matrix, 
    \begin{align*}
        \aa^\top \ggamma &= A_{\max}(\aa')^\top \ggamma \ge A_{\max}\|\AA'\|_{p \to q} - \AA_{\max}\varepsilon = \|\AA\|_{p \to q} - A_{\max}\varepsilon \\
        &\ge (1-\varepsilon)\|\AA\|_{p \to q}
    \end{align*}
    We can directly produce a stronger guarantee by extracting an approximate solution. Let $Y=\sum_j \yy(j)$ and $Z=\sum_i \zz(i)$, and define $\uu = \left(\frac{\yy}{Y}\right)^\frac{1}{p}$ and $\vv = \left(\frac{\zz}{Z}\right)^\frac{1}{q^*}$. We have $\|\uu\|_p^p = \|\vv\|_{q^*}^{q^*} = 1$, and using H\"older's inequality,
    \begin{align*}
        \aa^\top \ggamma &= \sum_{(i,j) \in S} \AA(i,j)\ggamma(i,j) \\
        &\le \sum_{(i,j) \in S} \AA(i,j)\vv(i)\uu(j) = \vv^\top \AA \uu \\
        &\le \|\vv\|_{q^*} \|\AA \uu\|_q = \|\AA \uu\|_q
    \end{align*}
    In particular, $\uu$ is a vector such that
    \begin{align*}
        \frac{\|\AA \uu\|_q}{\|\uu\|_p} \ge (1-\varepsilon)\|\AA\|_{p \to q} = (1-\varepsilon)\max_{\uu'} \frac{\|\AA \uu'\|_q}{\|\uu'\|_p} 
    \end{align*} \qedhere
\end{proof}

To turn Algorithm \ref{alg::IPM} into an efficient algorithm, we need to show that we can perform $\delta$-approximate Newton steps for the barrier function \eqref{eq::scf_barrier_F} quickly for constant accuracy $\delta < \frac{1}{5}$.

%% file: sections/05_newton.tex
\section{Fast Newton step using Laplacian solvers}\label{sec::newton}

In this section we will show how to efficiently compute $\delta$-approximate Newton steps for any fixed feasible point $\xx=(\zz, \yy, \ggamma) \in int(\mathcal{K})$. We can do it by computing the Schur complement of the Hessian matrix $\HH = \HH(\xx)$ and recognizing that it can be written as a symmetric M-matrix plus a rank-$2$ term, so it can be solved efficiently using Woodbury's identity and fast Laplacian solvers.

Let $\ww = (\zz,\yy)$ and recall the Hessian matrix \eqref{eq::hess_schur}
\begin{equation*}
    \HH = \begin{pmatrix}
        \HH_w & \HH_{w\gamma} \\
        \HH_{w\gamma}^\top & \HH_{\gamma} 
    \end{pmatrix}
\end{equation*}
Since $\HH_\gamma$ is diagonal, we reduce this system by computing the Schur complement \eqref{eq::schur-definition}
\begin{equation*}
    \HH\setminus \gamma = \HH_w - \HH_{w\gamma} \HH_\gamma^{-1} \HH_{w\gamma}^\top
\end{equation*}

\begin{theorem}\label{thm::hessian-schur}
    The Schur complement of the Hessian matrix of the self-concordant barrier \eqref{eq::scf_barrier_F} with respect to $\gamma$ can be written as
    \begin{equation*}
        \HH\setminus \gamma = \MM + \VV \VV^\top
    \end{equation*}
    Where $\MM \in \mathbb{R}^{(m+n)\times (m+n)} $ is a symmetric M-matrix with at most $4s$ non-zero entries, and $\VV \in \mathbb{R}^{(m+n) \times 2}$. Moreover, the matrix
    \begin{align*}
        \DD = \begin{pmatrix}
            Diag(\zz) & 0 \\
            0 & Diag(\yy)
        \end{pmatrix}
    \end{align*}
    has positive diagonal entries and makes $\DD \MM \DD$ strictly diagonally dominant.
\end{theorem}
\begin{proof}
    Define the following functions, for $a \ge 0, b \ge 0$ and $0 \le |c| < a^{\frac{1}{q^*}} b^\frac{1}{p}$.
    \begin{align*}
        \alpha(a,b,c) &= \frac{2}{q^*} \frac{a^\frac{2}{q^*} b^\frac{2}{p}}{a^\frac{2}{q^*} b^\frac{2}{p} - c^2}\left( 1 + \frac{2}{q^*}\frac{ c^2}{ a^\frac{2}{q^*}b^\frac{2}{p} - c^2} \right) + \frac{1}{q} 
        & \mu_z(a,b,c) = -\frac{4}{q^*}\frac{a^{\frac{2}{q^*}} b^\frac{2}{p} c}{(a^{\frac{2}{q^*}} b^\frac{2}{p} - c^2)^2}
        \\
        \alpha^*(a,b,c) &= \frac{2}{p} \frac{a^\frac{2}{q^*} b^\frac{2}{p}}{a^\frac{2}{q^*}b^\frac{2}{p} - c^2}\left( 1 + \frac{2}{p}\frac{ c^2}{ a^\frac{2}{q^*}b^\frac{2}{p} - c^2} \right) + \frac{1}{p^*} 
        & \mu_y(a,b,c) = -\frac{4}{p}\frac{a^{\frac{2}{q^*}} b^\frac{2}{p} c}{(a^{\frac{2}{q^*}} b^\frac{2}{p} - c^2)^2}  
        \\
        \beta(a,b,c) &= \frac{4}{pq^*} \frac{a^\frac{2}{q^*} b^\frac{2}{p} c^2}{( a^\frac{2}{q^*} b^\frac{2}{p} - c^2)^2} 
        & \mu_\gamma(a,b,c) = \frac{2(a^{\frac{2}{q^*}} b^\frac{2}{p} + c^2)}{(a^{\frac{2}{q^*}} b^\frac{2}{p} - c^2)^2}
    \end{align*}
    
    Using \eqref{eq::hessian_z2}-\eqref{eq::hessian_ycross}, note that $\HH_w$ has the following structure
    \begin{align*}
        \HH_w &= \begin{pmatrix}
            \HH_z & \HH_{z y} \\
            \HH_{z y}^\top & \HH_{y} 
        \end{pmatrix}
        =
        \begin{pmatrix}
            \DD_z + \frac{\mathbf{11}^\top}{(1-Z)^2} & \HH_{z y} \\
            \HH_{z y}^\top & \DD_y + \frac{\mathbf{11}^\top}{(1-Y)^2}
        \end{pmatrix}
        \\
        &=
        \begin{pmatrix}
            \DD_z & \HH_{z y} \\
            \HH_{z y}^\top & \DD_y
        \end{pmatrix}
        +
        \begin{pmatrix}
            \frac{\mathbf{11}^\top}{(1-Z)^2} & 0 \\
            0 & \frac{\mathbf{11}^\top}{(1-Y)^2}
        \end{pmatrix}
        \defeq \bar{\HH}_w + \VV \VV^\top
    \end{align*}
    where $\VV = (\frac{\mathbf{1}_m}{1-Z}, 0; 0, \frac{\mathbf{1}_n}{1-Y}) \in \mathbb{R}^{(m+n) \times 2}$, and $\DD_z$ and $\DD_y$ are diagonal matrices with entries 
    \begin{align*}
        \DD_z(i,i) &= \sum_{j : (i,j) \in S} \frac{\alpha(\zz(i),\yy(j), \ggamma(i,j))}{\zz(i)^2} 
        \\
        \DD_y(j,j) &= \sum_{i:(i,j) \in S} \frac{\alpha^*(\zz(i),\yy(j), \ggamma(i,j))}{\yy(j)^2} 
    \end{align*}
    The off-diagonal entries of $\bar{\HH}_w$ are computed in equation \eqref{eq::hessian_zy}, and they are also non-zero only when $(i,j) \in S$ with value 
    \begin{align*}
        \HH_{zy}(i,j) = \frac{\beta(\zz(i),\yy(j),\ggamma(i,j))}{\zz(i) \yy(j)}
    \end{align*}
    Consider the $2 \times 2$ contribution of edge $(i,j) \in S$ to $\bar{\HH}_w$
    \begin{align*}
        \HH_w^{(i,j)} = \begin{pmatrix}
            \frac{\alpha(\zz(i),\yy(j),\ggamma(i,j))}{\zz(i)^2} & \frac{\beta(\zz(i),\yy(j),\ggamma(i,j))}{\zz(i) \yy(j)} \\
            \frac{\beta(\zz(i),\yy(j),\ggamma(i,j))}{\zz(i)\yy(j)} & \frac{\alpha^*(\zz(i),\yy(j),\ggamma(i,j))}{\yy(j)^2}
        \end{pmatrix}
    \end{align*}
    If we identify this matrix with its $(m+n) \times (m+n)$ lift padded by zeroes, we can decompose the first term of $\HH_w$ as a sum over ``edges'' plus a rank-$2$ matrix.
    \begin{align*}
        \HH_w = \sum_{(i,j) \in S} \HH_w^{(i,j)} + \VV \VV^\top
    \end{align*}
   
    Consider now the Schur correction term $\HH_{w,\gamma} \HH_\gamma^{-1} \HH_{w,\gamma}^\top$. As discussed in Appendix \ref{sec::app_derivatives}, the matrix $\HH_{w\gamma}$ can be thought of as the unsigned edge-vertex incidence matrix of a bipartite graph with biadjacency matrix given by $\AA$ and weights given by \eqref{eq::hessian_zgam} and \eqref{eq::hessian_ygam}. Since $\HH_\gamma$ is diagonal, $\HH_\gamma^{-1}$ is simply re-scaling all the edge weights by \eqref{eq::hessian_gam2}. Therefore, we can define the matrix $\HH_{SC}^{(i,j)}$ using the functions $\mu_z, \mu_y, \mu_\gamma$, and write the Schur correction term as a sum over edges.
    \begin{align*}
        \HH_{SC}^{(i,j)} &\defeq \begin{pmatrix}
            \frac{\mu_z(\zz(i),\yy(j),\ggamma(i,j))^2}{\zz(i)^2\mu_\gamma(\zz(i),\yy(j),\ggamma(i,j))} & \frac{\mu_z(\zz(i),\yy(j),\ggamma(i,j))\mu_y(\zz(i),\yy(j),\ggamma(i,j))}{\zz(i)\yy(j) \mu_\gamma(\zz(i),\yy(j),\ggamma(i,j))} 
            \\
            \frac{\mu_z(\zz(i),\yy(j),\ggamma(i,j))\mu_y(\zz(i),\yy(j),\ggamma(i,j))}{\zz(i) \yy(j) \mu_\gamma(\zz(i),\yy(j),\ggamma(i,j))} & \frac{\mu_y(\zz(i),\yy(j),\ggamma(i,j))^2}{\yy(j)^2 \mu_\gamma(\zz(i),\yy(j),\ggamma(i,j))}
        \end{pmatrix}
        \\
        \HH_{w,\gamma} \HH_\gamma^{-1} \HH_{w,\gamma}^\top &= \sum_{(i,j) \in S} \HH_{SC}^{(i,j)}
    \end{align*}
    We can now write the full Schur complement as a sum over edge matrices plus a rank-$2$ matrix.
    \begin{align*}
        \HH \setminus \ggamma = \HH_w - \HH_{w,\gamma} \HH_\gamma^{-1} \HH_{w,\gamma}^\top = \sum_{(i,j) \in S} \left( \HH_w^{(i,j)} - \HH_{SC}^{(i,j)} \right) + \VV \VV^\top \defeq \sum_{(i,j) \in S} \MM^{(i,j)} + \VV \VV^\top
    \end{align*}
    We denote $\MM^{(i,j)}$ as the \textit{local edge matrix}, which is a $2 \times 2$ matrix (padded with zeroes). Letting $z = \zz(i), y = \yy(j)$, and $\gamma = \ggamma(i,j)$, we get
    \begin{align}\label{eq::local-edge-matrix}
        \MM^{(i,j)} &= \begin{pmatrix}
            \frac{\alpha(z,y,\gamma)}{z^2} - \frac{\mu_z(z,y,\gamma)^2}{z^2 \mu_\gamma(z,y,\gamma)} & \frac{\beta(z,y,\gamma)}{z y} - \frac{\mu_z(z,y,\gamma) \mu_y(z,y,\gamma)}{z y \mu_\gamma(z,y,\gamma)} 
            \\
            \frac{\beta(z,y,\gamma)}{z y} - \frac{\mu_z(z,y,\gamma)\mu_y(z,y,\gamma)}{z y \mu_\gamma(z,y,\gamma)} & \frac{\alpha^*(z,y,\gamma)}{y^2} - \frac{\mu_y(z,y,\gamma)^2}{y^2 \mu_\gamma(z,y,\gamma)}
        \end{pmatrix}
        \defeq
        \begin{pmatrix}
            \frac{1}{z^2}M_1 & \frac{1}{z y} M_{\text{cross}} \\
            \frac{1}{z y} M_{\text{cross}} & \frac{1}{y^2}M_2
        \end{pmatrix}
    \end{align}

    We can simplify the entries of this matrix by introducing the variable 
    \begin{align*}
        r = \frac{\gamma^2}{z^{\frac{2}{q^*}} y^{\frac{2}{p}}}. 
    \end{align*}
    Note that $0 \le r < 1$, and
    \begin{align*}
        \alpha(z,y,\gamma) &= \frac{4}{(q^*)^2} \frac{r}{(1-r)^2} + \frac{2}{q^*} \frac{1}{1-r} + \frac{1}{q} 
        & \frac{\mu_z(z,y,\gamma)^2}{\mu_\gamma(z,y,\gamma)} = \frac{4}{(q^*)^2}\frac{r}{1-r^2}\frac{2}{1-r} 
        \\
        \alpha^*(z,y,\gamma) &= \frac{4}{p^2} \frac{r}{(1-r)^2} + \frac{2}{p} \frac{1}{1-r} + \frac{1}{p^*} 
        & \frac{\mu_y(z,y,\gamma)^2}{\mu_\gamma(z,y,\gamma)} = \frac{4}{p^2}\frac{r}{1-r^2}\frac{2}{1-r} 
        \\
        \beta(z,y,\gamma) &= \frac{4}{p q^*}\frac{r}{(1-r)^2} 
        & \frac{\mu_z(z,y,\gamma)\mu_y(z,y,\gamma)}{\mu_\gamma(z,y,\gamma)} = \frac{4}{p q^*} \frac{r}{1-r^2}\frac{2}{1-r}
    \end{align*}
    We can use this to simplify the entries of \eqref{eq::local-edge-matrix} as follows.
    \begin{align*}
        M_1 &= \alpha(z,y,\gamma) - \frac{\mu_z(z,y,\gamma)^2}{\mu_\gamma(z,y,\gamma)} 
        = \frac{4}{(q^*)^2} \frac{r}{(1-r)^2} + \frac{2}{q^*} \frac{1}{1-r} + \frac{1}{q} - \frac{4}{(q^*)^2}\frac{r}{1-r^2}\frac{2}{1-r} \\
        &= \frac{4r}{(q^*)^2(1-r)^2} \left(1 - \frac{2}{1+r} \right) + \frac{2}{q^*} \frac{1}{1-r} + \frac{1}{q} \\
        &= -\frac{4}{(q^*)^2}\frac{r}{(1-r^2)} + \frac{1}{q^*} \frac{2}{1-r} + \frac{1}{q}
        \\
        M_2 &= \alpha^*(z,y,\gamma) - \frac{\mu_y(z,y,\gamma)^2}{\mu_\gamma(z,y,\gamma)} 
        = \frac{4}{p^2} \frac{r}{(1-r)^2} + \frac{1}{p} \frac{2}{1-r} + \frac{1}{p^*} - \frac{4}{p^2}\frac{r}{1-r^2}\frac{2}{1-r} \\
        &= \frac{4r}{p^2(1-r)^2} \left(1 - \frac{2}{1+r} \right) + \frac{1}{p} \frac{2}{1-r} + \frac{1}{p^*} \\
        &= -\frac{4}{p^2}\frac{r}{(1-r^2)} + \frac{1}{p} \frac{2}{1-r} + \frac{1}{p^*}
        \\
        M_{\text{cross}} &= \beta(z,y,\gamma) - \frac{\mu_z(z,y,\gamma)\mu_y(z,y,\gamma)}{\mu_\gamma(z,y,\gamma)}
        = \frac{4}{pq^*} \frac{r}{(1-r)^2} - \frac{4}{pq^*} \frac{r}{1-r^2}\frac{2}{1-r} \\
        &= \frac{4}{pq^*}\frac{r}{(1-r)^2} \left(1 - \frac{2}{1+r} \right) \\
        &= - \frac{4}{pq^*}\frac{r}{(1-r^2)} 
    \end{align*}
    We can simplify this even further by noting that $M_1 > -M_{\text{cross}}$ and $M_2 > -M_{\text{cross}}$.
    \begin{align*}
        M_1 &= - M_{\text{cross}} + M_1 + M_{\text{cross}} = - M_{\text{cross}} - \frac{4}{q^*} \frac{r}{1-r^2} \left(\frac{1}{p} + \frac{1}{q^*} \right) + \frac{2}{q^*(1-r)} + \frac{1}{q} 
        \\
        &= - M_{\text{cross}} + \frac{2}{q^*} \frac{1}{1-r} \left(1 - \frac{2r}{1+r} + \frac{2r}{1+r} \left(\frac{1}{q} - \frac{1}{p} \right)\right) + \frac{1}{q} \\
        &= - M_{\text{cross}} + \frac{2}{q^*} \frac{1}{1+r} \left( 1 + \frac{2r}{1-r} \left(\frac{1}{q} - \frac{1}{p} \right) \right) + \frac{1}{q}
        \\
        M_2 &= - M_{\text{cross}} + M_2 + M_{\text{cross}} = - M_{\text{cross}} - \frac{4}{p} \frac{r}{1-r^2} \left(\frac{1}{q^*} + \frac{1}{p} \right) + \frac{2}{p(1-r)} + \frac{1}{p^*} 
        \\
        &= - M_{\text{cross}} + \frac{2}{p} \frac{1}{1-r} \left(1 - \frac{2r}{1+r} + \frac{2r}{1+r} \left(\frac{1}{q} - \frac{1}{p} \right)\right) + \frac{1}{p^*} \\
        &= - M_{\text{cross}} + \frac{2}{p} \frac{1}{1+r} \left( 1 + \frac{2r}{1-r} \left(\frac{1}{q} - \frac{1}{p} \right) \right) + \frac{1}{p^*}
    \end{align*}
    
    Both remainders are strictly positive because $r < 1$ and $p \ge q$. To conclude the proof, let $\DD^{(i,j)} = Diag(z(i),y(j))$ padded with zeroes, and consider again its $2 \times 2$ restriction on a fixed edge where we drop the indices, then
    \begin{align*}
        \DD^{(i,j)}\MM^{(i,j)}\DD^{(i,j)} &= \begin{pmatrix}
            M_1 & M_{\text{cross}} \\
            M_{\text{cross}} & M_2
        \end{pmatrix}
        = 
        \frac{4r}{pq^*(1-r^2)}\begin{pmatrix}
            1 & -1 \\
            -1 & 1
        \end{pmatrix}
        \\
        &\hspace{1cm}
        +
        \begin{pmatrix}
            \frac{1}{q^*} \frac{2}{r+1} \left( 1 + \frac{2r}{1-r} \left(\frac{1}{q} - \frac{1}{p} \right) \right) + \frac{1}{q} & 0 \\
            0 & \frac{1}{p}\frac{2}{r+1} \left( 1 + \frac{2r}{1-r} \left(\frac{1}{q} - \frac{1}{p} \right) \right) + \frac{1}{p^*}
        \end{pmatrix}
    \end{align*}
    
    The first term is the Laplacian matrix of an edge with non-negative weight, and the second term is a diagonal matrix with positive diagonal entries, which implies that $\DD^{(i,j)}\MM^{(i,j)}\DD^{(i,j)}$ is a strictly symmetric diagonally dominant M-matrix, so each local edge matrix is an M-matrix. The matrix $\MM$ is the sum of the zero-padded edge matrices, and it has non-zero rows or columns because every vertex is incident to at least one edge. Therefore, $\MM$ is itself a symmetric M-matrix with scaling factor $\DD = Diag(\zz, \yy)$. 

    We can construct $\DD$ in linear time and each $\DD^{(i,j)}\MM^{(i,j)}\DD^{(i,j)}$ in constant time, so the matrix $\DD \MM \DD$ can be constructed in time $O(s)$, since it is a sum over edges. Each edge matrix has $4$ non-zero entries, so $\MM$ has at most $m+n+2s \le 4s$ non-zero entries.
\end{proof}

\subsection{The algorithm}\label{sec::fast-algo}

Let $0 < \delta < 1$. From Theorem \ref{thm::laplacian-solvers}, there is an algorithm \textsc{STSolve} which on input $\LL,\delta$ computes a program \textsc{solveL} such that, for every $\bb \in \mathbb{R}^n$, it returns $\textsc{solveL}(\bb) = \PP \bb$ for some symmetric matrix $\PP$ satisfying
\begin{align}\label{eq::laplacian-operatorbound}
    (1-\delta) \LL^{-1} \preceq \PP \preceq (1+\delta) \LL^{-1}
\end{align}

Letting $\widetilde{\xx} = \textsc{solveL}(\bb)$, this guarantees the following local norm bound
\begin{align}\label{eq::laplacian-error}
    \| \Tilde{\xx} - \LL^{-1}\bb\|_{\LL} \le \delta \|\LL^{-1} \bb\|_{\LL}
\end{align}
We can use this to construct an algorithm $\textsc{MSolve}$ with the same local norm approximation guarantee.

\begin{algorithm}[H]
\SetAlgoLined
\caption{\textsc{MSolve}}\label{alg::MSolve}
\SetKwInOut{Input}{Input}
\SetKwInOut{Output}{Output}
\Input{Laplacian solver \textsc{solveL}, diagonal matrix $\DD \in \mathbb{R}^{n \times n}$ with strictly positive diagonal entries, vector $\bb \in \mathbb{R}^n$.}
\Output{Vector $\widetilde{\xx} \in \mathbb{R}^n$ such that $\|\widetilde{\xx} - \MM^{-1}\bb \|_{\MM} \le \delta \|\MM^{-1} \bb\|_{\MM}$ whenever $\textsc{solveL} = \textsc{STSolve}(\DD \MM \DD,\delta)$}
$\widetilde{\uu} = \textsc{solveL}(\DD \bb)$ 
\\
$\widetilde{\xx} = \DD\widetilde{\uu}$
\\
\Return{$\widetilde{\xx}$}
\end{algorithm}

We can now present an algorithm for solving systems of the form $(\MM + \VV \VV^\top)\xx = \bb$, as the Schur complement of the Hessian of the barrier \ref{eq::scf_barrier_F} from Theorem \ref{thm::hessian-schur}. This is based on Woodbury identity, i.e.
\begin{equation}\label{eq::schur-woodbury}
    (\MM + \VV \VV^\top)^{-1} = \MM^{-1} - \MM^{-1}\VV(I + \VV^\top \MM^{-1} \VV)^{-1} \VV^\top \MM^{-1}
\end{equation}

\begin{algorithm}[H]
\SetAlgoLined
\caption{\textsc{SchurSolve}}\label{alg::SchurSolve}
\SetKwInOut{Input}{Input}
\SetKwInOut{Output}{Output}
\Input{Positive definite SDD matrix $\LL \in \mathbb{R}^{n \times n}$ with $s$ non-zero entries, diagonal matrix $\DD \in \mathbb{R}^{n \times n}$ with strictly positive diagonal entries, matrix $\VV = (\vv_1, \vv_2) \in \mathbb{R}^{n\times 2}$, accuracy parameter $0 < \delta < 1$, vector $\bb \in \mathbb{R}^n$}
\Output{Vector $\widetilde{\xx} \in \mathbb{R}^n$ such that $\|\widetilde{\xx} - \AA^{-1}\bb \|_{\AA} \le \delta \|\AA^{-1}\bb\|_{\AA}$, where $\AA=\DD^{-1} \LL \DD^{-1}+\VV \VV^\top$}
$\textsc{solveL} = \textsc{STSolve}(\LL,\delta)$
\\
$\widetilde{\xx}_1 = \textsc{MSolve}(\textsc{solveL}, \DD, \bb)$ (Algorithm \ref{alg::MSolve})
\\
$\widetilde{\uu}_1 = \textsc{MSolve}(\textsc{solveL}, \DD, \vv_1)$, $\widetilde{\uu}_2 = \textsc{MSolve}(\textsc{solveL},\DD,\vv_2)$, $\widetilde{\UU}=(\widetilde{\uu}_1, \widetilde{\uu}_2)$
\\
$\widetilde{\xx}_2 = \VV^\top \widetilde{\xx}_1$ 
\\
Solve the $2 \times 2$ system $(I+\VV^\top \widetilde{\UU})\widetilde{\xx}_3 = \widetilde{\xx}_2$
\\
$\widetilde{\xx} = \widetilde{\xx}_1 - \widetilde{\UU} \widetilde{\xx}_3$
\\
\Return{$\widetilde{\xx}$}
\end{algorithm}

Algorithm \ref{alg::SchurSolve} runs in total time $\widetilde{O}(s \log\frac{1}{\delta})$, which is dominated by construction of the approximate Laplacian solver and its three applications in steps $1$-$3$. Note that these are three calls of the solver for the same $L$ and $\delta$, so they apply the same linear operator to the different right-hand sides. The additional work coming from the $\VV \VV^\top$ term is in step $5$, which is a $2 \times 2$ linear system that can be formed in $O(s)$ time and solved in constant time. If we replace the applications of $\textsc{MSolve}$ with exact $\MM^{-1}$, we obtain an exact solution to the system, which follows from substituting into \eqref{eq::schur-woodbury}.

We can use Algorithm \ref{alg::SchurSolve} to compute the full Newton step
\begin{align*}
    \begin{pmatrix}
        \HH_w & \HH_{w\gamma} \\
        \HH_{w\gamma}^\top & \HH_\gamma
    \end{pmatrix}
    \begin{pmatrix}
        \DDelta_w \\
        \DDelta_\gamma
    \end{pmatrix}
    =
    -\begin{pmatrix}
        \gg_w \\
        \gg_\gamma
    \end{pmatrix}
\end{align*}

\begin{algorithm}[H]
\SetAlgoLined
\caption{\textsc{NewtonSolve}}\label{alg::NewtonSolve}
\SetKwInOut{Input}{Input}
\SetKwInOut{Output}{Output}
\Input{Matrix $\AA \in \mathbb{R}^{m \times n}$ with $s$ non-zero entries, accuracy parameter $\delta > 0$, point $\xx=(\zz,\yy,\ggamma) \in int(\mathcal{K})$, central path parameter $\eta > 0$, exponents $p \ge q > 1$.}
\Output{Vector $\widetilde{\DDelta} \in \mathbb{R}^{m+n+s}$ such that $\|\widetilde{\DDelta} - \DDelta \|_{\xx} \le \delta \|\DDelta\|_x$, where $\DDelta = -\HH(\xx)^{-1} \nabla \Phi_\eta(\xx)$.}
$p^* = \frac{p}{p-1}$, $q^* = \frac{q}{q-1}$
\\
$Z = \sum_{i=1}^m \zz(i)$, $Y = \sum_{j=1}^n \yy(j)$
\\
$\rr(i,j) = \frac{\ggamma(i,j)^2}{\zz(i)^\frac{2}{q^*} \yy(j)^\frac{2}{p}}$, $\rr_1(i,j) = \frac{4\rr(i,j)}{pq^*(1-\rr(i,j)^2)}$, $\rr_2(i,j) = \frac{2}{1+\rr(i,j)}\left( 1 + \frac{2\rr(i,j)}{1-\rr(i,j)} \left(\frac{1}{q} - \frac{1}{p}\right) \right)$
\\
Compute the gradient $(\gg_z, \gg_y, \gg_\gamma) = \nabla F(\xx) -  \eta \bar{\aa}$ using \eqref{eq::grad_z} - \eqref{eq::grad_gam}, denote $\gg_w = (\gg_z, \gg_y)$.
\\
Compute the Hessian sub-matrices $\HH_{w,\gamma}$ and $\HH_{\gamma}$ using \eqref{eq::hessian_gam2} - \eqref{eq::hessian_ygam}.
\\
Compute the following matrices, where the $\LL^{(i,j)}$ matrices are padded with zeros.
\begin{align*}
    \DD &= \begin{pmatrix}
        Diag(\zz) & 0 \\
        0 & Diag(\yy)
    \end{pmatrix}
    &
    \VV = \begin{pmatrix}
        \frac{\mathbf{1}_m}{1-Z} & 0 \\ 
        0 & \frac{\mathbf{1}_n}{1-Y}
    \end{pmatrix}
    \\
    \LL^{(i,j)} &=  \rr_1(i,j)\begin{pmatrix}
        1 & -1 \\
        -1 & 1
    \end{pmatrix}
    +
    \begin{pmatrix}
        \frac{1}{q^*}\rr_2(i,j) + \frac{1}{q} & 0 \\
        0 & \frac{1}{p}\rr_2(i,j) + \frac{1}{p^*}
    \end{pmatrix}
    &
    \LL = \sum_{(i,j) \in S}\LL^{(i,j)}
\end{align*}
\\
$\widetilde{\gg}_w = \gg_w - \HH_{w\gamma} \HH_{\gamma}^{-1}\gg_\gamma$
\\
$\widetilde{\DDelta}_w = \textsc{SchurSolve}(\LL,\DD,\VV,\delta,-\widetilde{\gg}_w)$ (Algorithm \ref{alg::SchurSolve})
\\
$\tilde{\gg}_\gamma = \gg_\gamma + \HH_{w\gamma}^\top \widetilde{\DDelta}_w$
\\
$\widetilde{\DDelta}_\gamma = -\HH_\gamma^{-1} \tilde{\gg}_\gamma$
\\
\Return{$\widetilde{\DDelta} = (\widetilde{\DDelta}_w, \widetilde{\DDelta}_\gamma) $}
\end{algorithm}

We will use Algorithm \ref{alg::NewtonSolve} as the key IPM subroutine for computing approximate Newton steps in Algorithm \ref{alg::IPM}. The following Lemma establishes the desired approximation guarantees.

\begin{lemma}\label{lem::newtonsolve-works}
    Let $\xx=(\zz,\yy,\ggamma) \in int(\mathcal{K})$, $\HH = \nabla^2 F(\xx)$, $0 < \delta < 1$ and $\eta > 0$. Let $\DDelta = -\HH^{-1}\nabla \Phi_\eta(\xx)$ be the Newton step and $\widetilde{\DDelta}$ be the output of Algorithm \ref{alg::NewtonSolve}, then
    \begin{align*}
        \|\widetilde{\DDelta} - \DDelta\|_{\HH} \le \delta \|\DDelta\|_{\HH}
    \end{align*}
    Moreover, $\widetilde{\DDelta}$ is computed in time $\widetilde{O}(s \log\frac{1}{\delta})$.
\end{lemma}
\begin{proof}
    We begin by observing that one way to solve a system of equations $\HH \DDelta = \bb$ is to compute the Schur complement $\HH \setminus \ggamma$ and solve
    \begin{align}
        (\HH\setminus \gamma)\DDelta_w &= \bb_w - \HH_{w\gamma} \HH_\gamma^{-1}\bb_\gamma \label{eq::newton-schur-w}\\
        \HH_\gamma \DDelta_\gamma &= \bb_\gamma - \HH_{w\gamma}^\top \DDelta_w \label{eq::newton-schur-gam}
    \end{align}
    The Newton step is a special case of this where $\bb = -\nabla \Phi_\eta(\xx)$, but our analysis works for general $\bb \in \mathbb{R}^{m+n+s}$. Theorem \ref{thm::hessian-schur} establishes the decomposition
    \begin{align*}
        \BB = \HH\setminus \gamma = \MM + \VV \VV^\top
    \end{align*}
    Therefore, Algorithm \ref{alg::NewtonSolve} is implementing an approximate solution $\widetilde{\DDelta}$ to $\DDelta$ by solving \eqref{eq::newton-schur-w} and \eqref{eq::newton-schur-gam} applying Woodbury's identity to \eqref{eq::newton-schur-w} and replacing applications of exact inverses $\MM^{-1}$ with the approximate version given by \textsc{MSolve}. More precisely, recall that \textsc{solveL} implicitly constructs a matrix $\PP$ satisfying \eqref{eq::laplacian-operatorbound}, so $\textsc{MSolve}$ (Algorithm \ref{alg::MSolve}) is equivalent to applying a matrix $\PP_1$ such that
    \begin{align*}
        (1-\delta) \MM^{-1} \preceq &\PP_1 \preceq (1+\delta)\MM^{-1} \\
        \frac{1}{1+\delta} \MM \preceq &\PP_1^{-1} \preceq \frac{1}{1-\delta} \MM
    \end{align*}
    On the other hand, \textsc{SchurSolve} (Algorithm \ref{alg::SchurSolve}) is implementing Woodbury's identity to solve a system for $\BB = \MM + \VV \VV^\top$ replacing $\MM^{-1}$ with $\PP_1$, so it applies the operator $(\PP_1^{-1} + \VV \VV^\top)^{-1}$ to approximate $\BB^{-1}$. Therefore,
    \begin{align*}
        \frac{1}{1+\delta} \MM + \VV \VV^\top \preceq \PP_1^{-1} &+ \VV \VV^\top \preceq\frac{1}{1-\delta}\MM + \VV \VV^\top 
        \\
        \frac{1}{1+\delta} (\MM + \VV \VV^\top) \preceq \PP_1^{-1} &+ \VV \VV^\top \preceq\frac{1}{1-\delta} (\MM + \VV \VV^\top) 
        \\
        (1-\delta)\BB^{-1} = (1-\delta) (\MM + \VV \VV^\top)^{-1} \preceq (\PP_1^{-1} &+ \VV \VV^\top)^{-1} \preceq (1+\delta) (\MM + \VV \VV^\top)^{-1} =(1+\delta)\BB^{-1} 
        \\
        (1-\delta)I \preceq \BB^{\frac{1}{2}} (\PP_1^{-1} &+ \VV \VV^\top)^{-1} \BB^{\frac{1}{2}} \preceq (1+\delta)I
    \end{align*}
    
    This lets us approximate the approximation error in the $A$-norm with the same error incurred by the inner Laplacian solver. Let $\widetilde{\DDelta}_w = \textsc{SchurSolve}(\LL,\DD,\VV,\delta,\widetilde{\bb}_w) = (\PP_1^{-1} + \VV \VV^\top)^{-1} \widetilde{\bb}_w$, then 
    \begin{align*}
        \|\widetilde{\DDelta}_w - \DDelta_w\|_{\BB} &= \|(\PP_1^{-1} + \VV \VV^\top)^{-1} \widetilde{\bb}_w - \BB^{-1}\widetilde{\bb}_w\|_{\BB} \\
        &= \|(\BB^{\frac{1}{2}}(\PP_1^{-1} + \VV \VV^\top)^{-1}\BB^{\frac{1}{2}} - I)\BB^{-\frac{1}{2}} \widetilde{\bb}_w\|_2 \\
        &\le \delta \|\BB^{-\frac{1}{2}}\widetilde{\bb}_w\|_2 
        = \delta \|\BB^{-1}\widetilde{\bb}_w\|_{\BB} 
        = \delta \|\DDelta_w\|_{\BB}
    \end{align*}

    We now lift this approximation error from the $\BB$-norm to the $\HH$-norm. This follows from the variational characterization of Schur complements. Setting $\xx_1 = \widetilde{\DDelta}_w - \DDelta_w$ and $\xx_2 = -\HH_\gamma^{-1} \HH_{w\gamma}^\top (\widetilde{\DDelta}_w - \DDelta_w) = (\widetilde{\DDelta}_\gamma - \DDelta_\gamma)$ in Lemma \ref{lem::variational-schur} implies
    \begin{align*}
        \|(\widetilde{\DDelta}_w - \DDelta_w, \widetilde{\DDelta}_\gamma - \DDelta_\gamma)\|_{\HH} 
        = \|\widetilde{\DDelta}_w - \DDelta_w\|_{\BB}
        \le \delta \|\DDelta_w\|_{\BB}
    \end{align*}
    To conclude the proof, we complete the following square.
    \begin{equation*}
        \|(\DDelta_w, \DDelta_\gamma)\|_{\HH}^2 = \|\DDelta_w\|_{\BB}^2 + \|\DDelta_\gamma + \HH_\gamma^{-1}\HH_{w\gamma}^\top \DDelta_w\|_{\HH_\gamma}^2
    \end{equation*}
    In particular,
    \begin{align*}
        \|(\widetilde{\DDelta}_w - \DDelta_w, \widetilde{\DDelta}_\gamma - \DDelta_\gamma)\|_{\HH} &\le \delta \|\DDelta_w\|_{\AA} \le \delta \|(\DDelta_w, \DDelta_\gamma)\|_{\HH}
    \end{align*}

    The runtime analysis follows from the fact that steps $1$ through $7$ and $9-10$ can be performed in time $O(s)$, and invoking \textsc{SchurSolve} requires time $\widetilde{O}(s \log \frac{1}{\delta})$.
\end{proof}

Typically, Laplacian solvers are used as algorithms producing approximate solutions $\tilde{\xx}$ on inputs $\LL,\delta,\bb$ with the $\LL$-norm approximation property 
\begin{align*}
    \|\tilde{x} - \LL^{-1}\bb\|_{\LL} \le \delta \|\LL^{-1}\bb\|_{\LL}
\end{align*}
This is clearly weaker than Theorem \ref{thm::laplacian-solvers}, however it would be a natural starting point in the analysis since it suggests the natural proof strategy for Lemma \ref{lem::newtonsolve-works} of ``lifting'' the approximation property from $\LL$ to $\MM$ to $\BB$ to $\HH$. While the first and last step work directly, the second step does not work if we allow for errors to depend on the right-hand sides of the linear systems. Therefore, using the stronger form of Laplacian solvers as fixed linear operators is necessary for our analysis.

\bigskip

We can now conclude the proof of the main theorem.
\begin{proof}[Proof of Theorem \ref{thm::main}]
    If $q=1$, we can compute the norm exactly in $O(s)$ time because it has the following closed form.
    \begin{align*}
        \|\AA\|_{p \to 1} = \|\AA^\top \mathbf{1}\|_{p^*}
    \end{align*}
    Therefore, we can assume $p \ge q > 1$, which is exactly the regime where Algorithm \ref{alg::IPM} applies. By Lemma \ref{lem::termination}, after $T\le O(\sqrt{s} \log\frac{s}{\varepsilon})$ Algorithm \ref{alg::IPM} computes a point $\uu \in \mathbb{R}^n$ such that 
    \begin{align*}
        \frac{\|\AA \uu\|_{q}}{\|\uu\|_p} \ge (1 - \varepsilon)\|\AA\|_{p \to q}
    \end{align*}
    Initialization can be performed in $O(s)$ time (Lemma \ref{lem::analytic-center}), and each iteration requires $\widetilde{O}(s)$ time (Lemma \ref{lem::newtonsolve-works} with $\delta = \Theta(1)$). Therefore, the total algorithm runtime is $\widetilde{O}(s^{\frac{3}{2}} \log \frac{1}{\varepsilon})$.
\end{proof}

%% file: sections/06_experiments.tex
\section{Experiments}\label{sec::experiments}


We implemented our algorithm in Julia to compare its performance against state of the art solvers. We applied a few modifications to the algorithm to make it practical, in particular we used a long step IPM approach as described in \cite[Section 11.3]{boyd2004convex}. The algorithm is still a primal-only path -following IPM, where we use the dual only to construct certificates to monitor the duality gap for deciding when to stop. Two other notable modifications we made for practical purposes are replacing the simplex constraints in \eqref{eq::norm-conic} with equality constraints and using projected Newton steps, and we used the cheapest version of the Laplacian solver of \cite{GKS} which only builds the approximate Cholesky preconditioner without any Chebyshev iterations.

\begin{figure}[h]
    \centering
    \includegraphics[width=\linewidth]{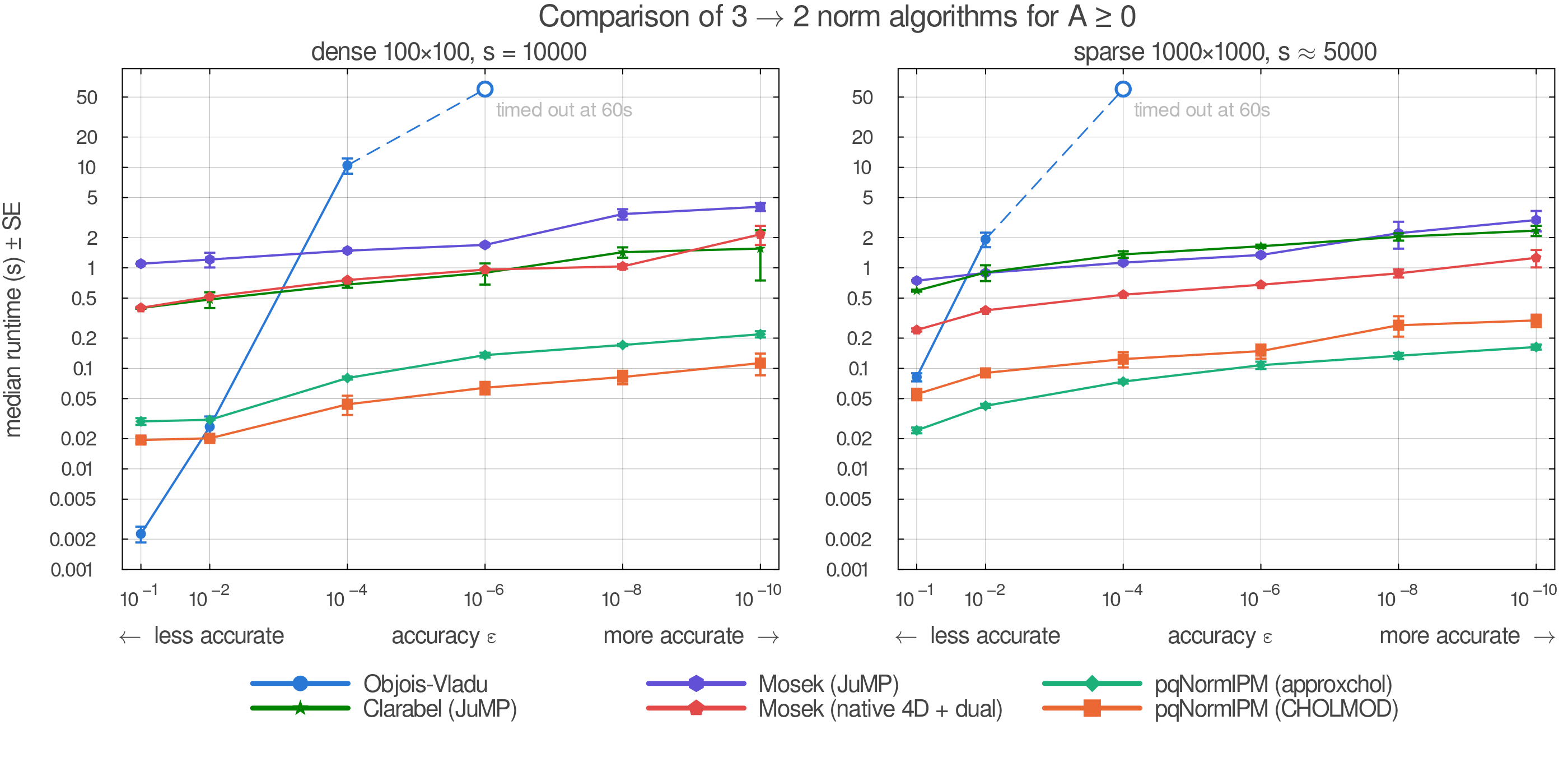}
    \caption{Runtime of different algorithms to compute $\|A\|_{3 \to 2}$ for different target accuracies, median over $5$ random matrices. \textbf{(Left).} Dense matrices with random entries in $[0,1]$ of size $100 \times 100$. \textbf{(Right).} Sparse binary matrices of size $1000 \times 1000$ where each entry is sampled iid with probablity $4/1000$, and the diagonal is set to $1$, so the average sparsity is highly concentrated around $5000$.}
    \label{fig::comparison}
\end{figure}

We compared our algorithm with an implementation of \cite{objois2026approximating},\footnote{Written using Claude.} together with two standard conic solvers (Clarabel \cite{Clarabel_2024} and MOSEK \cite{aps2024mosek}) available in Julia applied to the same program as our IPM \eqref{eq::norm-conic}. We also tuned the Mosek implementation by having it optimize the dual of \eqref{eq::norm-conic} and use the native 4-dimensional power cone for the case $p > q$.\footnote{Since JuMP and MOI only support the $3$-dimensional power cone and decompose $4$-dimensional cones using 3D ones, this was implemented directly with the Mosek Julia API, as opposed to MosekTools.} In this section, we focus on $p=3$ and $q=2$, and show similar results for $(p,q) \in \{(2,1.5), (3,3)\}$ in Appendix \ref{sec::experiments-app}.

\subsection{Hardware and software details}

All experiments were ran on an Apple Macbook Air (2022) with M2 CPU (8 cores, 4 performance and 4 efficiency) and 16GB of memory. We used Julia 11.1.2 with the following packages: Laplacians 1.4.1, SparseArrays 1.11.0, CHOLMOD 5.2.0, OpenBLAS 0.3.27, JuMP 1.24.0, MathOptInterface 1.36.0. The conic programming solvers we used are Clarabel.jl 0.11.0 and Mosek 11.0.22, called through Mosek.jl 11.0.1 and MosekTools.jl 0.15.10.

\subsection{Results}

Our algorithm is significantly faster than the baselines on both dense and sparse matrices, as shown in Figure \ref{fig::comparison}, across different levels of accuracy. The algorithm from \cite{objois2026approximating} is faster for low-accuracy solutions on dense matrices, but it quickly becomes too slow at moderate accuracy, due to the $1/\varepsilon$ dependence on runtime. We implemented two versions of our algorithm: one solving the Schur complement of the Hessian using the approximate Cholesky solver of \cite{GKS}, and one using optimized standard software for sparse Cholesky decomposition (CHOLMOD \cite{chen2008algorithm}). The former is the closest match to the algorithm analyzed in the paper (Algorithm \ref{alg::NewtonSolve}), and it is the fastest algorithm for sparse matrices, where it beats the optimized general solver.

\begin{figure}[h]
    \centering
    \includegraphics[width=\linewidth]{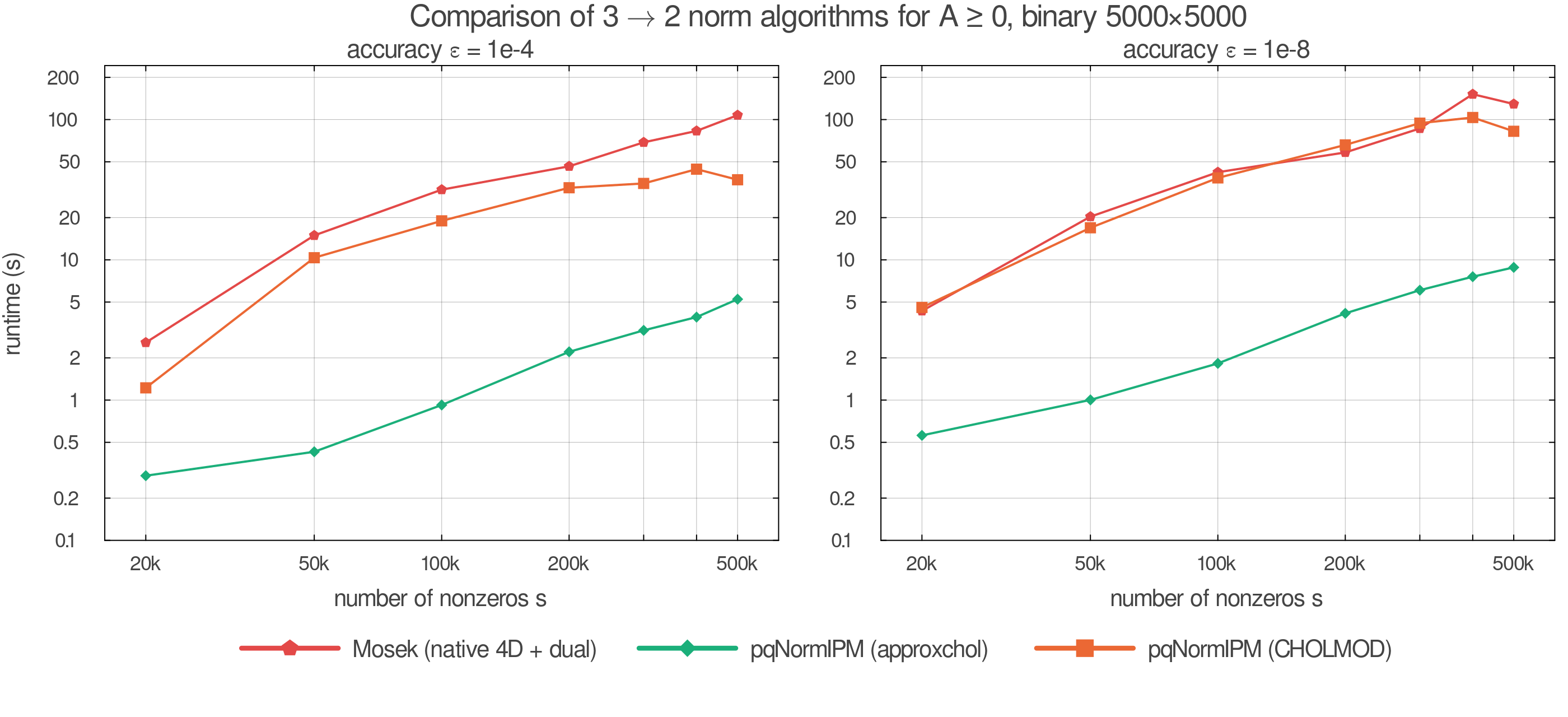}
    \caption{Performance of different algorithms for computing $3 \to 2$ norms of large scale non-negative matrices. Each point represents a single random matrix sampled at that sparsity level. \textbf{(Left.)} Medium accuracy solutions $\varepsilon=10^{-4}$. \textbf{(Right.)} High accuracy solutions $\varepsilon=10^{-8}$.}
    \label{fig::largescale}
\end{figure}

We also investigated the performance of our algorithm on larger scale matrices, which is represented in Figure \ref{fig::largescale}. Algorithm \ref{alg::IPM} is even stronger here, computing high accuracy norms in $20\times$ ($23\times$) faster than CHOLMOD (MOSEK) at $s=10^5$ and $9.3\times$ ($14.6\times$) faster than CHOLMOD (MOSEK) at $s=5\times 10^5$. At these scales, these differences are non-trivial.

The improvement between CHOLMOD and MOSEK is mostly explained by leveraging the structure of the program \eqref{eq::norm-conic} and performing a Schur complement, which solves a much smaller system. The second source of improvement in our algorithm is the fast Laplacian solver, which makes a significant difference on large matrices.

\medskip

Our implementation of Algorithm \ref{alg::IPM} is available at the following GitHub repository.
\begin{center}
\href{https://github.com/marcopirazzini/MatrixNormIPM.jl}{https://github.com/marcopirazzini/MatrixNormIPM.jl}
\end{center}

%% file: sections/A_IPMapp.tex
\section{Details on Interior Point Methods}\label{sec::app_IPM}

The specific IPM used to prove Theorem \ref{thm::IPM_full} corresponds to a primal short-step path following IPM, which was introduced by \cite{nesterov1994interior} and simplified by \cite{renegar2001mathematical} under exact Newton steps. The inexact Newton step analysis closely follows the one of \cite{li2017inexact}.
\begin{theorem}\label{thm::IPM_master}
    Let $\xx_0 \in int(\mathcal{K})$ and $\eta_0$ satisfy $\lambda_{\eta_0}(\xx_0) \le \frac{1}{10}$, and let $0 \le \delta \le \frac{1}{5}$. Let $\widetilde{\Delta}_\eta(\xx)$ be an approximate Newton step satisfying the local norm guarantee \eqref{eq::newton-approx}, and iteratively update $\xx$ and $\eta$ as follows. 
    \begin{align*}
        \xx_{+} &= \xx + \widetilde{\Delta}_\eta (\xx) \\
        \eta_{+} &= \left( 1 + \frac{1}{20\sqrt{\nu}}\right)\eta
    \end{align*}
    Then $\lambda_{\eta_+}(\xx_+) \le \frac{1}{10}$. In particular, we reach an approximate solution to $\xx_{\eta_T}^*$ after $T \le \left\lceil 21 \sqrt{\nu} \log\frac{\eta_T}{\eta_0}\right\rceil$ iterations.
\end{theorem}

We will prove Theorem \ref{thm::IPM_master} using two standard lemmas. The first auxiliary Lemma is a simplification of \cite[Theorem 2]{li2017inexact} when working with pure Newton steps and no penalty function $h(x)=0$.
\begin{lemma}[Inexact Newton step]\label{lem::newton}
    Let $F:\mathcal{K} \to \mathbb{R}$ be a self-concordant function and $\xx \in int(\mathcal{K})$ be a point such that $\lambda(\xx) = \lambda < 1$. If $\widetilde{\DDelta}$ approximates the Newton step as $\|\Tilde{\DDelta} - \Delta(\xx)\|_{\xx} \le \delta \|\Delta(\xx)\|_{\xx}$ for some $\delta < \min \{ \frac{1}{2}, \frac{1-\lambda}{\lambda} \}$, then $\xx_+ = \xx + \Tilde{\DDelta}$ satisfies
    \begin{equation*}
        \lambda(\xx_+) \le \frac{\delta \lambda + (1+\delta) \lambda^2}{(1 - (1+\delta) \lambda)^2}
    \end{equation*}
\end{lemma}

The second auxiliary Lemma is an elementary result from \cite[Eq. 2.4.5]{renegar2001mathematical}.
\begin{lemma}[Path update step]\label{lem::IPM_path}
    Let $F$ be a self-concordant barrier function with complexity parameter $\nu$, and let $\xx \in int(\mathcal{K})$. For any $\eta > 0$ and $\alpha > 0$, a multiplicative scaling of the barrier parameter $\eta_+ = (1+\alpha)\eta$ satisfies
    \begin{equation*}
        \lambda_{\eta_+}(\xx) \le (1 + \alpha) \lambda_\eta (\xx) + \alpha \sqrt{\nu}
    \end{equation*}
\end{lemma}

\begin{proof}[Proof of Theorem \ref{thm::IPM_master}]
    Let $\lambda = \lambda_\eta(\xx) \le \frac{1}{10}$ and $\alpha = \frac{1}{20\sqrt{\nu}}$. We bound $\lambda_{\eta_+}(\xx_+)$ as a function of $\lambda_{\eta}(\xx_+)$ using the path update step (Lemma \ref{lem::IPM_path}), and then bound $\lambda_{\eta}(\xx_+)$ as a function of $\lambda_{\eta}(\xx)$ using the Newton step (Lemma \ref{lem::newton}).
    \begin{align*}
        \lambda_{\eta_+}(\xx_+) &\le (1+\alpha)\lambda_\eta(\xx_+) + \alpha \sqrt{\nu} \\
        &\le (1+\alpha)\frac{\delta \lambda + (1+\delta) \lambda^2}{(1 - (1+\delta) \lambda)^2} + \alpha \sqrt{\nu} \\
        &\le \frac{21}{20} \frac{\frac{4}{125}}{\left(\frac{22}{25}\right)^2} + \frac{1}{20} \\
        &= \frac{21}{484} + \frac{1}{20} < \frac{1}{10}
    \end{align*}
    The bound on the number of iterations immediately follows from 
    \begin{align*}
        \eta_T = \left(1 + \frac{1}{20\sqrt{\nu}} \right)^T \eta_0 &\iff T = \frac{\log\frac{\eta_T}{\eta_0}}{\log\left({1+\frac{1}{20\sqrt{\nu}}}\right)}
    \end{align*}
    Since $\log(1+\frac{1}{x}) \ge \frac{1}{x+1}$ for any $x \ge 1$, a sufficient condition for a large enough such $T$ is
    \begin{equation*}
        T = \left\lceil 21\sqrt{\nu}\log\frac{\eta_T}{\eta_0} \right\rceil
    \end{equation*}
\end{proof}
The final auxiliary Lemma controls how close an approximate minimizer for $\Phi_\eta (\xx)$ is to the minimizer of the constrained linear problem as a function of $\eta$. Recall a useful property of self-concordant functions.
\begin{enumerate}
    \item[\textbf{(SC3)}] For any $\xx,\yy \in int(\mathcal{K})$ with $\|\xx-\yy\|_{\xx} < 1$, we can swap the reference point for the local norm and still get a bound on their distance,
    \begin{equation*}
        \|\xx-\yy\|_{\yy} \le \frac{\|\xx-\yy\|_{\xx}}{1-\|\xx-\yy\|_{\xx}}
    \end{equation*}
    \label{itm::sc3}
\end{enumerate}

\begin{lemma}[Termination]\label{lem::IPM_termination}
    For any $\xx \in int(\mathcal{K})$ such that $\lambda_{\eta}(\xx) \le \frac{1}{10}$, we have
    \begin{equation*}
        \langle \cc, \xx\rangle - \langle \cc, \xx^*\rangle \le \frac{7}{6}\frac{\nu}{\eta}
    \end{equation*}
\end{lemma}
\begin{proof}
    This is again a standard result in path-following IPMs adapted to our settings. Eq. (2.4.4) from \cite{renegar2001mathematical} states that, if $\| \xx - \xx_\eta^*\|_{\xx_\eta^*} \le \beta$, then
    \begin{equation*}
        \langle \cc, \xx\rangle - \langle \cc, \xx^*\rangle \le (1+ \beta) \frac{\nu}{\eta}
    \end{equation*}
    Proposition 2.2.8 from \cite{renegar2001mathematical} implies that
    \begin{equation*}
        \|\xx - \xx_\eta^*\|_{\xx} \le \lambda_\eta(\xx) + \frac{3\lambda_\eta(\xx)^2}{(1-\lambda_\eta(\xx))^3} \le \frac{1}{7}
    \end{equation*}
    We can use (\hyperref[itm::sc3]{SC3}) to change the reference point of the local norm,
    \begin{equation*}
        \| \xx - \xx_\eta^*\|_{\xx_\eta^*} \le \frac{\| \xx - \xx_\eta^*\|_{\xx}}{1 - \| \xx - \xx_\eta^*\|_{\xx}} \le \frac{1}{6}
    \end{equation*}
    which concludes the proof.
\end{proof}

\medskip

We can combine the IPM inductive theorem and the termination lemma to prove Theorem \ref{thm::IPM_full}.

\begin{proof}[Proof of Theorem \ref{thm::IPM_full}]
    From Theorem \ref{thm::IPM_master}, we need $\left\lceil 21 \sqrt{\nu} \log \frac{\eta_T}{\eta_0}\right\rceil$ iterations to reach a point $\xx_T$ such that $\lambda_{\eta_T}(\xx_T) \le \frac{1}{10}$. Set $\varepsilon= \frac{7\nu}{6\eta_T}$, then Lemma \ref{lem::IPM_termination} implies that 
    \begin{equation*}
        \langle \cc, \xx_T\rangle - \langle \cc, \xx^*\rangle \le \varepsilon
    \end{equation*}
    This means that the number of iterations needed is
    \begin{equation*}
        T \le \left\lceil 21 \sqrt{\nu} \log \frac{7\nu}{6\varepsilon \eta_0} \right\rceil \le \left\lceil 22 \sqrt{\nu} \log \frac{\nu}{\varepsilon \eta_0} \right\rceil
    \end{equation*}
    The final inequality is valid for any $0 < \eta_0 < 1$, $0 < \varepsilon < 1$ and large enough $\nu \ge 40$.
\end{proof}

%% file: sections/B_derivatives.tex
\section{Gradient and Hessian of the barrier function}\label{sec::app_derivatives}

In this section we collect the formulas for the gradient and Hessian of the self-concordant barrier function \eqref{eq::scf_barrier_F}. To help readability, we recall the explicit form of the barrier here. Let $Z = \sum_i \zz(i)$ and $Y = \sum_j \yy(j)$,
\begin{align*}
    F(\zz,\yy,\ggamma) = - \sum_{(i,j) \in S} \left( \log \left(\zz(i)^\frac{2}{q^*} \yy(j)^\frac{2}{p} - \ggamma(i,j)^2\right) + \frac{1}{q}\log \zz(i) + \frac{1}{p^*} \log \yy(j)\right)  \\
    - \log(1-Z) - \log(1-Y) 
\end{align*}
For two indices $1 \le i \le m$ and $1 \le j \le n$, if $(i,j) \in S$ we say $i \sim j$ and $j \sim i$. 

\subsection{Gradient}
The gradients with respect to the first two vectors are very similar.
\begin{align}
    \frac{\partial F}{\partial \zz(i)} &= - \sum_{j \sim i} \frac{2}{q^*} \frac{\zz(i)^{\frac{2}{q^*}-1} \yy(j)^\frac{2}{p}}{ \zz(i)^\frac{2}{q^*} \yy(j)^\frac{2}{p} - \ggamma(i,j)^2} - \frac{d_i}{q \zz(i)} + \frac{1}{1-Z} \label{eq::grad_z} \\
    \frac{\partial F}{\partial \yy(j)} &= - \sum_{i \sim j} \frac{2}{p} \frac{\zz(i)^{\frac{2}{q^*}} \yy(j)^{\frac{2}{p} -1}}{\zz(i)^\frac{2}{q^*} \yy(j)^\frac{2}{p} - \ggamma(i,j)^2} - \frac{e_j}{p^* \yy(j)} + \frac{1}{1-Y} \label{eq::grad_y}\\
    \frac{\partial F}{\partial \ggamma(i,j)} &= \frac{2 \ggamma(i,j)}{\zz(i)^\frac{2}{q^*} \yy(j)^\frac{2}{p} - \ggamma(i,j)^2} \label{eq::grad_gam}
\end{align}

\subsection{Hessian}
The Hessian of the barrier function is highly structured. First, note that the gradient of the edge variables $\ggamma(i,j)$ \eqref{eq::grad_gam} depends only on $\ggamma(i,j)$ and not $\ggamma(i',j')$ for $(i',j') \ne (i,j)$, so the Hessian restricted to $\ggamma$ is diagonal. Moreover, \eqref{eq::grad_gam} only depends on the row/column variables corresponding to the edge $(i,j)$, so its derivative with respect to $\zz(i')$ or $\yy(j')$ for $i' \ne i$ and $j' \ne j$ is zero. The non-zero Hessian entries are as follows.

\begin{align}
    \frac{\partial^2 F}{\partial\zz(i)^2} &=  \sum_{j \sim i} \frac{2}{q^* \zz(i)^2} \frac{\zz(i)^\frac{2}{q^*} \yy(j)^\frac{2}{p}}{\zz(i)^\frac{2}{q^*}\yy(j)^\frac{2}{p} - \ggamma(i,j)^2}\left( 1 + \frac{2}{q^*}\frac{ \ggamma(i,j)^2}{ \zz(i)^\frac{2}{q^*}\yy(j)^\frac{2}{p} - \ggamma(i,j)^2} \right) \label{eq::hessian_z2} \\
    &+ \frac{d_i}{q \zz(i)^2} + \frac{1}{(1-Z)^2}  \nonumber \\ 
    \frac{\partial^2 F}{\partial \zz(i)\partial \zz(i')} &=  \frac{1}{(1-Z)^2} \qquad (i \ne i') \label{eq::hessian_zcross} \\
    \frac{\partial^2 F}{\partial \yy(j)^2} &= \sum_{i \sim j} \frac{2}{p\yy(j)^2} \frac{\zz(i)^\frac{2}{q^*} \yy(j)^\frac{2}{p}}{\zz(i)^\frac{2}{q^*}\yy(j)^\frac{2}{p} - \ggamma(i,j)^2}\left( 1 + \frac{2}{p}\frac{ \ggamma(i,j)^2}{ \zz(i)^\frac{2}{q^*}\yy(j)^\frac{2}{p} - \ggamma(i,j)^2} \right) \label{eq::hessian_y2} \\
    &+ \frac{e_j}{p^* \yy(j)^2} + \frac{1}{(1-Y)^2} \nonumber \\
    \frac{\partial^2 F}{\partial \yy(j)\partial \yy(j')} &= \frac{1}{(1-Y)^2} \qquad (j \ne j') \label{eq::hessian_ycross} \\
    \frac{\partial^2 F}{\partial \zz(i) \partial \yy(j)} &= \frac{4}{pq^* \zz(i)\yy(j)} \frac{\zz(i)^\frac{2}{q^*} \yy(j)^\frac{2}{p} \ggamma(i,j)^2}{( \zz(i)^\frac{2}{q^*} y(j)^\frac{2}{p} - \ggamma(i,j)^2)^2} \cdot \mathbbm{1}\left\{(i,j) \in S\right\}\label{eq::hessian_zy} \\
    \frac{\partial^2 F}{\partial \ggamma(i,j)^2} &= \frac{2(\zz(i)^\frac{2}{q^*} \yy(j)^\frac{2}{p} + \ggamma(i,j)^2)}{(\zz(i)^\frac{2}{q^*} \yy(j)^\frac{2}{p} - \ggamma(i,j)^2)^2} \label{eq::hessian_gam2} \\
    \frac{\partial^2 F}{\partial \ggamma(i,j) \partial \zz(i)} &= -\frac{4}{q^*\zz(i)} \frac{\zz(i)^\frac{2}{q^*} \yy(j)^\frac{2}{p} \ggamma(i,j)}{(\zz(i)^\frac{2}{q^*} \yy(j)^\frac{2}{p} - \ggamma(i,j)^2)^2}  \label{eq::hessian_zgam} \\
    \frac{\partial^2 F}{\partial \ggamma(i,j) \partial \yy(j)} &= -\frac{4}{p\yy(j)} \frac{\zz(i)^\frac{2}{q^*} \yy(j)^\frac{2}{p} \ggamma(i,j)}{(\zz(i)^\frac{2}{q^*} \yy(j)^\frac{2}{p} - \ggamma(i,j)^2)^2} \label{eq::hessian_ygam} 
\end{align}

The Hessian matrix can be represented as follows, where the rows and columns are indexed according to $\xx=(\zz,\yy,\ggamma)$.
\begin{equation*}
    \HH = \begin{pmatrix}
        \HH_z & \HH_{zy} & \HH_{z\gamma} \\
        \HH_{zy}^\top & \HH_{y} & \HH_{y \gamma} \\
        \HH_{z\gamma}^\top & \HH_{y \gamma}^\top & \HH_{\gamma}
    \end{pmatrix}
\end{equation*}

There is an even simpler representation of the Hessian by grouping the row and column variables into a single vector $\ww=(\zz,\yy)$ and focusing on $\ggamma$.
\begin{equation}\label{eq::hess_schur}
    \HH = \begin{pmatrix}
        \HH_w & \HH_{w\gamma} \\
        \HH_{w\gamma}^\top & \HH_{\gamma} 
    \end{pmatrix}
\end{equation}
where
\begin{align*}
    &\HH_w = \begin{pmatrix}
        \HH_z & \HH_{zy} \\
        \HH_{zy}^\top & \HH_{y}
    \end{pmatrix} 
    &
    \HH_{w\gamma} = \begin{pmatrix}
        \HH_{z\gamma} \\
        \HH_{y \gamma}
    \end{pmatrix}
\end{align*}
This way of looking at the Hessian is particularly useful because $\HH_\gamma$ is diagonal, so solving systems of equations in $\HH$ reduces to solving systems of equations in the Schur complement of $\HH$ with respect to $\HH_\gamma$. Also note that every column of $\HH_{w\gamma}$ is supported on two entries, one in the $\zz$ block and one in the $\yy$ block. In particular, this can be thought of as the unsigned edge-vertex incidence matrix of a bipartite graph with incidence matrix given by $\AA$ and vertex weights given by \eqref{eq::hessian_zgam} and \eqref{eq::hessian_ygam}.

%% file: sections/C_experiments_app.tex
\section{Experiments on different matrix norms}\label{sec::experiments-app}

In this section, we report the performance of our algorithms for $(p,q) \in \{(2,1.5),(3,3)\}$ on experiments analogous to Section \ref{sec::experiments}. This demonstrates that our algorithm remains practical for a wide range of norms, and it beats current alternatives by a wide margin.

\begin{figure}[h]
    \centering
    \includegraphics[width=0.79\linewidth]{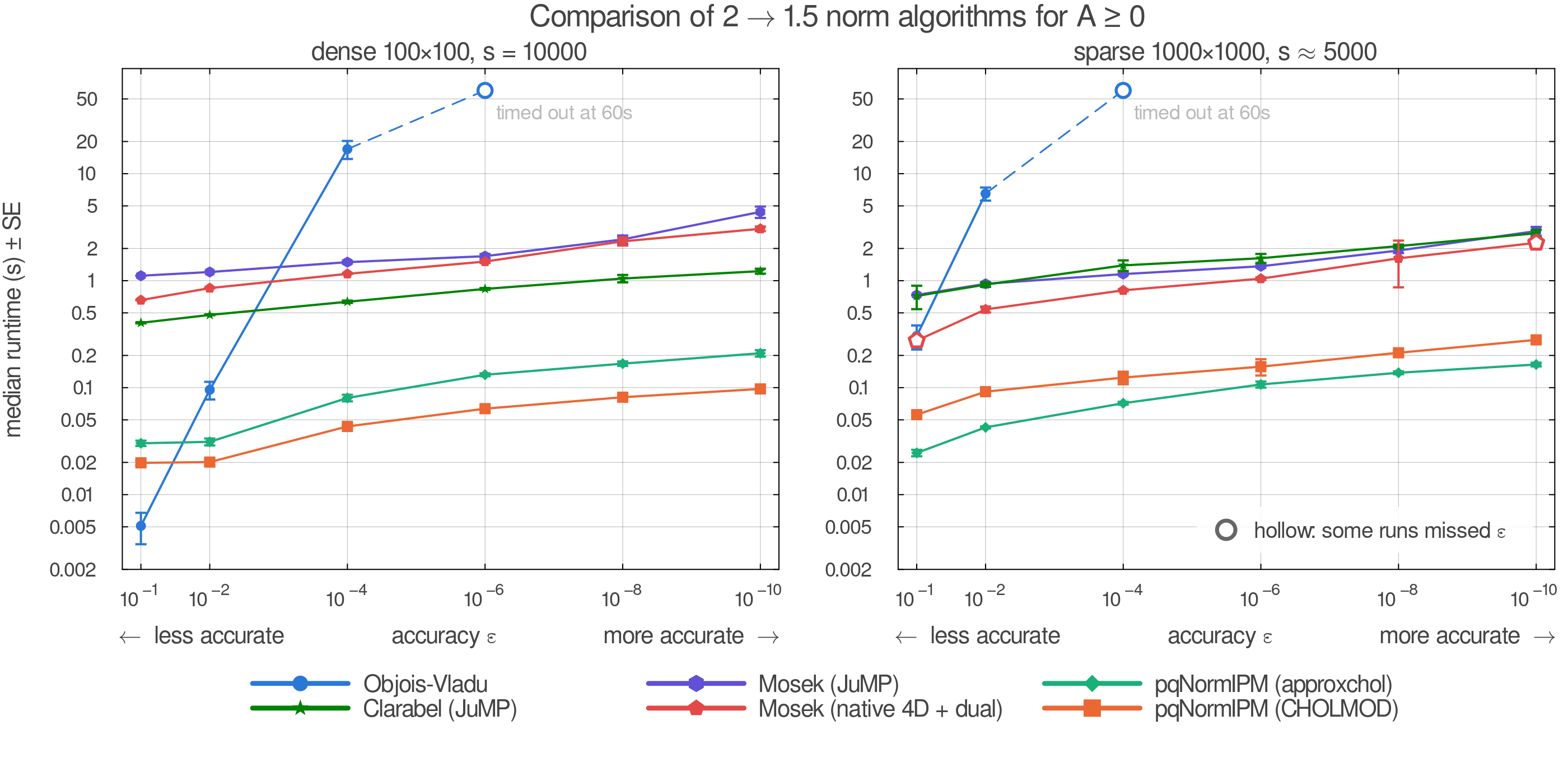}
    \caption{Runtime of different algorithms to compute $\|A\|_{2 \to 1.5}$ for different target accuracies, median over $5$ random matrices. \textbf{(Left).} Dense matrices with random entries in $[0,1]$ of size $100 \times 100$. \textbf{(Right).} Sparse binary matrices of size $1000 \times 1000$ where each entry is sampled iid with probablity $4/1000$, and the diagonal is set to $1$, so the average sparsity is highly concentrated around $5000$.}
    \label{fig::comparison_2_15}
\end{figure}

\begin{figure}[h]
    \centering
    \includegraphics[width=0.79\linewidth]{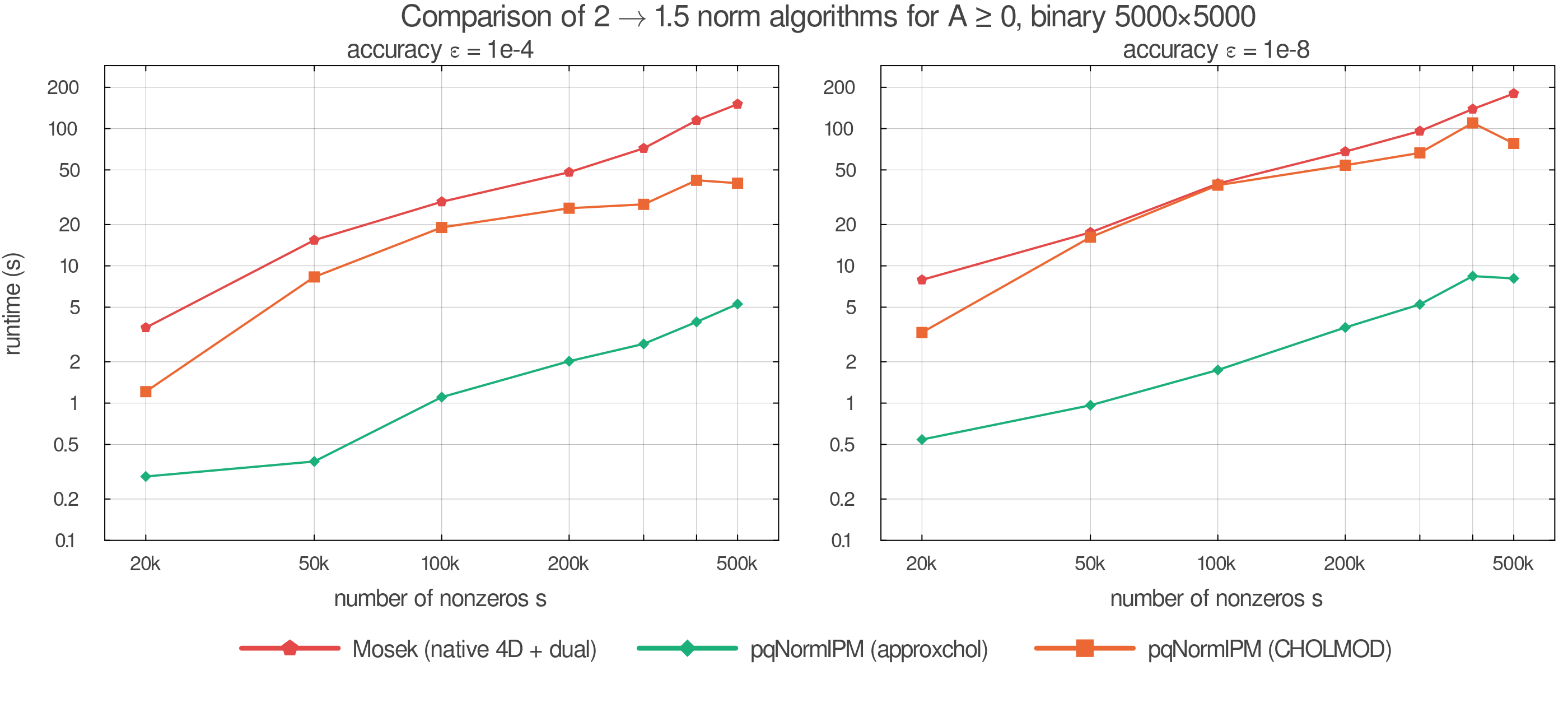}
    \caption{Performance of different algorithms for computing $2 \to 1.5$ norms of large scale non-negative matrices. Each point represents a single random matrix sampled at that sparsity level. \textbf{(Left.)} Medium accuracy solutions $\varepsilon=10^{-4}$. \textbf{(Right.)} High accuracy solutions $\varepsilon=10^{-8}$.}
    \label{fig::largescale_2_15}
\end{figure}

\newpage

\begin{figure}[h]
    \centering
    \includegraphics[width=0.79\linewidth]{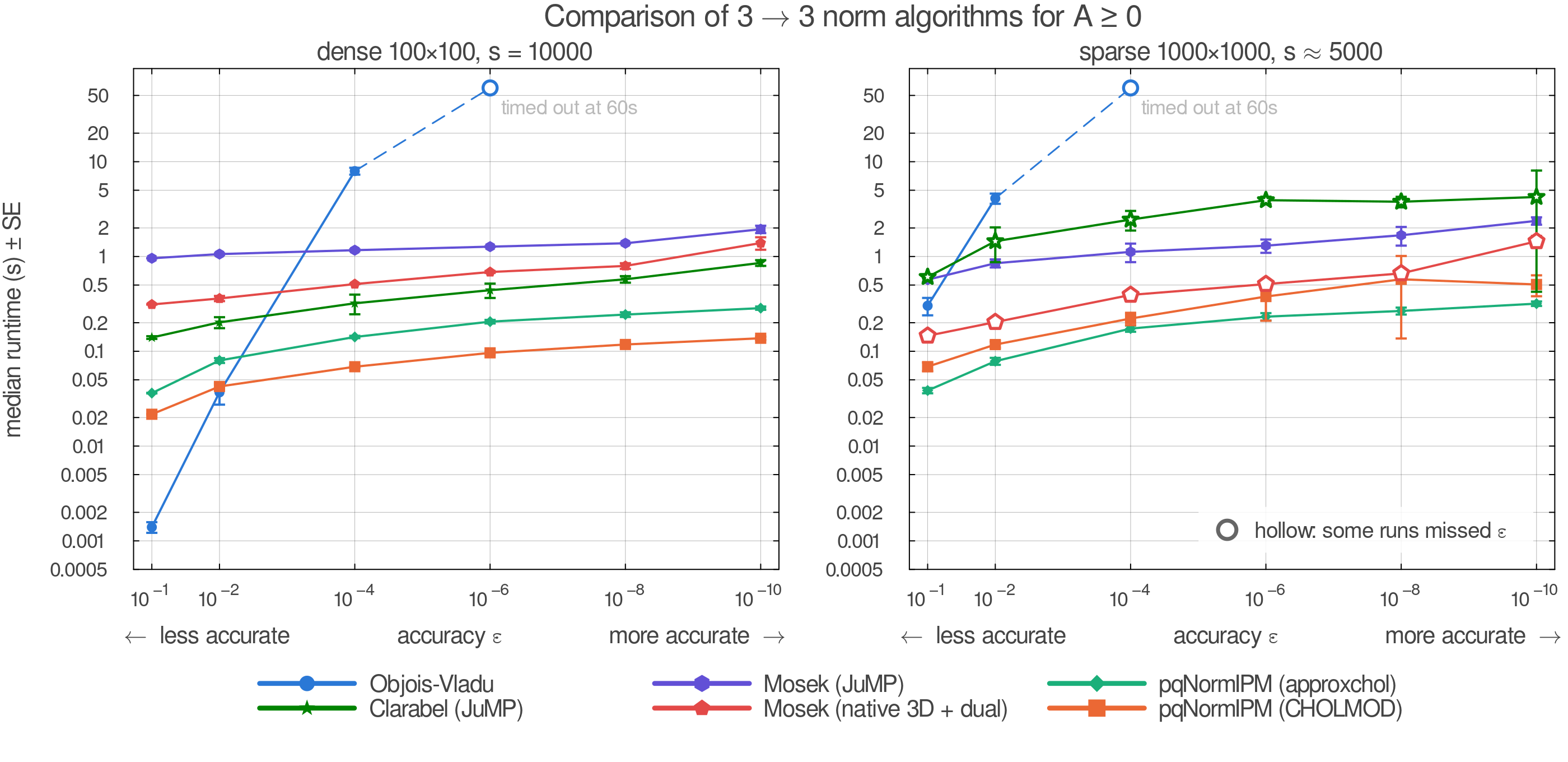}
    \caption{Runtime of different algorithms to compute $\|A\|_{3 \to 3}$ for different target accuracies, median over $5$ random matrices. \textbf{(Left).} Dense matrices with random entries in $[0,1]$ of size $100 \times 100$. \textbf{(Right).} Sparse binary matrices of size $1000 \times 1000$ where each entry is sampled iid with probablity $4/1000$, and the diagonal is set to $1$, so the average sparsity is highly concentrated around $5000$.}
    \label{fig::comparison_3_3}
\end{figure}

\begin{figure}[h]
    \centering
    \includegraphics[width=0.79\linewidth]{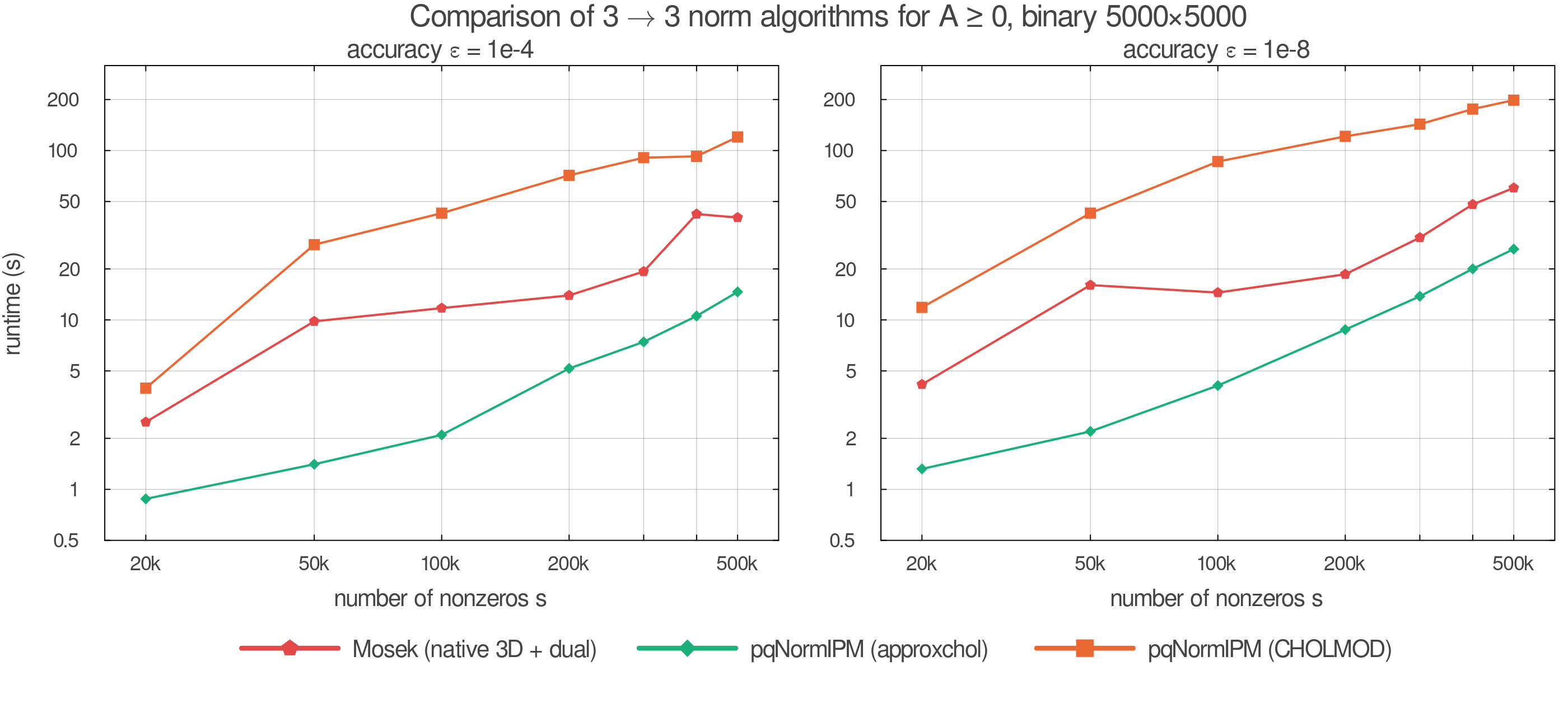}
    \caption{Performance of different algorithms for computing $3 \to 3$ norms of large scale non-negative matrices. Each point represents a single random matrix sampled at that sparsity level. \textbf{(Left.)} Medium accuracy solutions $\varepsilon=10^{-4}$. \textbf{(Right.)} High accuracy solutions $\varepsilon=10^{-8}$.}
    \label{fig::largescale_3_3}
\end{figure}

As witnessed in Figure \ref{fig::comparison_3_3}, some implementations are fragile when $p=q$, and fail to reach convergence. Our algorithm is slower when $p=q$ compared to $p > q$, but it always converges to the target accuracy, and it still comfortably outperforms the other baselines. This might be due to the differences in the $3$D and $4$D power cones, and it would be interesting to investigate this further.